\documentclass[fleqn,usenatbib,useAMS]{mnras}
\usepackage{fix-cm}
\newcommand{\kms}{\mbox{\,km s}^{-1}}
\newcommand{\Msun}{ M_{\odot} }

\newcommand{\pcc}{{\mbox{\,cm}^{-3}}}
\newcommand{\K}{ \rm K}
\newcommand{\ppcc}{{\rm cm}^{-3}}

\newcommand{\eff}{\epsilon_{\rm ff}}

\newcommand{\tff}{t_{\rm ff}}

\newcommand{\Msinks}{M_{\rm sinks}}
\newcommand{\SFR}{{\rm SFR}}
\newcommand{\SFE}{{\rm SFE}}
\newcommand{\Sigmagas}{\Sigma_{\rm gas}}
\newcommand{\Mgas}{M_{\rm gas}}

\newcommand{\Mmol}{M_\mathrm{mol}}

\newcommand{\Msf}{M_{\rm VDG}}

\newcommand{\Sigmamol}{\Sigma_{\rm mol}}
\newcommand{\Sigmasfr}{\Sigma_{\rm {SFR}}}

\newcommand{\taudepl}{\tau_{\rm depl}}
\newcommand{\tauff}{{\tau_{\rm ff}}}

\newcommand{\tausf}{{\tau_{\mathrm{YSO}}}}
\newcommand{\tausfr}{{\tau_e}}
\newcommand{\dotm}{{\dot{M}_*}}

\newcommand{\blue} {\color{blue} }

\def\alamenos#1{$^{-#1}$}

\def\promedio#1{\langle{#1}\rangle}
\def\average#1{\langle{#1}\rangle}

\def\apj{ApJ}
\def\apjl{ApJL}
\def\aap{A\& A}

\def\araa{ARA\&A}
\def\mnras{MNRAS}

\def\aj{{AJ}}
\def\apjs{ApJS}

\def\mnras{{MNRAS}}
\def\pasj{{PASJ}}
\def\pasp{{PASP}}

\usepackage{graphicx}	% Including figure files
\usepackage{amsmath}	% Advanced maths commands
\usepackage{amssymb}	% Extra maths symbols
\usepackage{multicol}        % Multi-column entries in tables
\usepackage{bm}		% Bold maths symbols, including upright Greek
\usepackage{pdflscape}	% Landscape pages

\usepackage{xcolor}
\usepackage{ulem}

\usepackage[T1]{fontenc}
\usepackage{ae,aecompl}

\usepackage{newtxtext,newtxmath}
\title[Schmidt-Kennicutt relations in collapsing clouds]{Gravity or turbulence? VII. 
The Schmidt-Kennicutt law, the star formation efficiency, and the mass density of clusters from gravitational collapse rather than turbulent support
}

\author[Zamora-Avilés et al.]{Manuel Zamora-Avilés$^{1}$,
Vianey Camacho$^{2,3}$,
Javier Ballesteros-Paredes$^2$\thanks{Contact e-mail: \href{j.ballesteros@irya.unam.mx}{j.ballesteros@irya.unam.mx}}, 
Enrique Vázquez-Semadeni$^2$, 
\newauthor
Aina Palau$^2$, Carlos Román-Zúñiga$^4$, 
Andrés Hernández-Cruz$^1$, Gilberto C. Gómez$^2$, Fabián
\newauthor
Quesada-Zúñiga$^{1}$, and Raúl Naranjo-Romero$^5$
\\
$^{1}$Instituto Nacional de Astrof\'isica, \'Optica y Electr\'onica, Luis E. Erro 1, Tonantzintla, 72840 Puebla, M\'exico\\
$^{2}$Instituto de Radioastronom\'ia y Astrof\'isica, UNAM, Apartado Postal 3-72, 58089 Morelia Michoac\'an, M\'exico\\
$^{3}$Center of Astronomy and Gravitation, Department of Earth Sciences, National Taiwan Normal University, 88, Sec. 4, Ting-Chou Rd., \\
Wenshan District, Taipei 116, Taiwan R.O.C\\
$^4$Instituto de Astronom\'ia, Universidad Nacional Aut{\'o}noma de M{\'e}xico, Unidad Acad{\'e}mica en Ensenada, Ensenada 22860 BC, M{\'e}xico \\
$^5$InnovaBienestar de M{\'e}xico, SAPI de CV. Ciencia y Tecnolog{\'i}a 790,  Saltillo 400, 25290, Coahuila de Zaragoza, M{\'e}xico. \\
}

\date{Last updated 2020 June 10; in original form 2013 September 5}

\pubyear{2023}

\usepackage[utf8]{inputenc}

\begin{document}
\label{firstpage}
\pagerange{\pageref{firstpage}--\pageref{lastpage}}
\maketitle

% Abstract of the paper
\begin{abstract}

We explore the Schmidt-Kennicutt (SK) relations and the star formation efficiency per free-fall time ($\epsilon_{\rm ff}$), mirroring observational studies, in numerical simulations of filamentary molecular clouds undergoing gravitational contraction. We find that (a)~collapsing clouds accurately replicate the observed SK relations for galactic clouds and (b)~$\epsilon_{\rm ff}$ is small and constant in space and in time, with values similar to those found in local clouds. We propose that this constancy arises from the similar radial scaling of the free-fall time ($\tau_{\rm ff}$) and the internal mass in density structures with spherically-averaged density profiles near $r^{-2}$. We additionally show that (c)~the star formation rate (SFR) increases rapidly in time; (d)~the low values of $\epsilon_{\rm ff}$ result from evaluating $\tauff$ and the characteristic star-formation time scale over different time intervals, combined with the increasing SFR, and (e)~the fact that star clusters are significantly denser than the gas clumps from which they form is a natural consequence of the rapidly increasing SFR, the continuous replenishment of the star-forming gas by the accretion flow, and the near $r^{-2}$ density profile induced by the collapse. Finally, we argue that interpreting $\epsilon_{\rm ff}$ as an efficiency is problematic since it is not bounded by unity, and because the gas mass in clouds evolves. Instead, we propose that viewing $\epsilon_{\rm ff}$ as the ratio of the actual SFR to the  gas free-fall rate. In summary, our results show that the SK relation, the low values of $\epsilon_{\rm ff}$, and the mass density of stellar clusters arise naturally from gravitational contraction.

\end{abstract}

% Select between one and six entries from the list of approved keywords.
% Don't make up new ones.
\begin{keywords}
galaxies: star formation -- ISM: clouds -- Galaxy: fundamental parameters
\end{keywords}

%%%%%%%%%%%%%%%%%%%%%%%%%%%%%%%%%%%%%%%%%%%%%%%%%%

%%%%%%%%%%%%%%%%% BODY OF PAPER %%%%%%%%%%%%%%%%%%

\section{Introduction} \label{sec:intro}

The Schmidt-Kennicutt (SK) relation is a fundamental correlation in astrophysics that has important implications for our understanding of how galaxies form stars and evolve. In its classical form is an empirical relation between the surface densities of the star formation rate ($\Sigmasfr$) and the gas mass ($\Sigmagas$). It was first proposed by \citet{Schmidt59} and confirmed observationally by \citet{Kennicutt98}. The relation has a power law form $\Sigmasfr$~$\propto$~$\Sigmagas^N$, where the exponent $N$ is typically between 1 and 2 for extragalactic studies \citep[see the review by][and references therein]{Kennicutt_Evans12, Sun+23}, but around 2 or larger for individual molecular clouds \citep[e.g.,][]{Gutermuth+11, Pokhrel+21}.

Here, we focus on individual molecular cloud (MC) observational determinations of the SK relation with the more reliable determination of the star formation rate, based on direct young stellar objects (YSO) counts \citep[see, e.g.,][]{Gutermuth+11, Lada+13, Evans+14, Lombardi+14, Heyer+16, Zari+16, Ochsendorf+17, Lada+17, Pokhrel+21}.\footnote{Other works based on indirect determinations of the SFR give SK relation with a wide dispersion \citep[e.g.,][]{Lee+16, Gallagher+18, Leroy+25}.} These observations produce a classical $\Sigmasfr\propto\Sigmagas^N$ SK relation with an exponent $N$~$\sim$~2 with some dispersion. \\

{ From the very definition of the star formation rate, SFR$\equiv dM_*/dt \equiv \dotm$, where $M_*$ is the mass in stars in a certain region, one can derive a fundamental star formation law \citep[see, e.g., ][and references therein for a formal derivation]{BP+23},}
\begin{equation}
  \dotm = \eff \bigg(\frac{\Mgas}{\tauff}\bigg),
  \label{eq:fundamental}
  \end{equation}
where $\Mgas$ is the mass in gas, $\eff$ a parameter known as the {\it star formation efficiency per free-fall time}, and $\tauff$ is the free-fall time given by
\begin{equation}
  \tauff=\sqrt{\frac{3\pi}{32 G \rho}},
  \label{eq:tauff}
\end{equation}
with $\rho$ the mean mass density of the system. On the other hand, we can write $\eff$ as
\begin{equation} %\label{eq:eff}
 % \eff = \frac{\tauff}{\taudepl} .
  \eff =  \frac{\dotm}{\Mgas / \tauff} = \frac{\tauff}{\taudepl},
  \label{eq:eff}
\end{equation}
with $\taudepl$ being the depletion time, which is the time to exhaust the {current} gas mass, given the {current} star formation rate $\dotm = dM_*/dt$,
\begin{equation} \label{eq:taudepl}
  \taudepl = \frac{\Mgas}{\dotm}. \\
\end{equation}

Equation (\ref{eq:fundamental}) has been proposed based on theoretical and empirical arguments elsewhere \citep[e.g., ][]{Krumholz_Tan07, Krumholz+12, Lee+16, Vutisalchavakul+16, Elmegreen_2018, Pokhrel+21, Sun+23}. { Traditionally, $\eff$ is called the star formation efficiency per free-fall time. Here, we just prefer to refer to it as the {\it ratio of the SFR to the gas-infall rate}, since, as we will discuss in \S\ref{sec:problems_eff}, its interpretation as an efficiency is problematic. For clarity, we keep its traditional symbol, $\eff$, and quantify it following standard observational approaches \citep[e.g.,][see \ref{subsec:structure}]{Pokhrel+21}.}\footnote{{ 
We emphasize that our intention is to reinterpret $\eff$, not to redefine it, as we discuss in Sec.\ \ref{sec:problems_eff}.}}

Observationally, the star formation rate of a molecular cloud in the Solar Neighbourhood is approximated as the amount of mass present in some type of protostar, divided by the characteristic lifetime of those protostars, $\tausf$, 
\begin{equation}
  \promedio{\SFR}_\tausf = 
   \frac{M_{*}} {\tausf}.
   \label{eq:mean_SFR}
\end{equation}
Observations in the last decade towards Milky Way molecular clouds have reported values of $\eff$ in the range of 0.006-0.06 \citep[e.g.,][]{Lada+13, Evans+14, Heyer+16, Ochsendorf+17}. More recently, \citet{Pokhrel+21} {measured the SK relation in MCs within the Solar Neighbourhood. They used} complete counts of protostellar objects to estimate SFR and Herschel observations to estimate the surface density {at various column-density levels in the clouds. They found} that the surface density of star formation rate correlates, on the one hand, linearly with the ratio of the { molecular gas} surface density ($\Sigmamol$) divided by its free-fall time,
\begin{equation}
  \Sigmasfr \propto \frac{\Sigmamol}{\tauff},
  \label{eq:Pokhrel1}
\end{equation}
but on the other, quadratically with the surface density of the gas alone,
\begin{equation}
  \Sigmasfr \propto \Sigma^2_{\rm mol}.
  \label{eq:Pokhrel2}
\end{equation}
In addition, they also showed that the ratio of the SFR to the gas-infall rate ($\eff$), measured for different surface density thresholds within a single cloud, is nearly independent of the surface density, and that the set of clouds exhibits values of $\eff$ in the range 
\begin{equation}
  \eff \in (0.1, 0.01),
  \label{eq:Pokhrel3}
\end{equation}
with a mean value of $\eff\sim 0.03$. It is important to note that this value is derived at the scale of MCs due to resolution limitations. It is worth noting that the modified version of the SK law in terms of the $\Sigma_{\rm mol}/\tauff$ \citep[eq. \ref{eq:Pokhrel1};][]{Genzel+10, Krumholz+12}, can be viewed in general as a two-parameter relation of the form
\begin{equation}
    \log \Sigma_{\rm SFR} = \log \eff + N_2 \log \left(\frac{\Sigmamol} {\tauff}\right),
    \label{eq:SK_log}
\end{equation}
with $N_2$ the exponent ($N_2 =1$ in eq. \ref{eq:Pokhrel1}) and $\eff$ the intercept.

The long depletion times implied by the ratio $\Sigmagas/\Sigmasfr$ \citep[$\sim10^9$~yr for galaxies, and $\gtrsim 10^7$ for local molecular clouds, see][]{Kennicutt98, Bigiel+08, Pokhrel+21} have been used to argue that star formation is a slow process \citep[e.g., ][]{Krumholz_Tan07, Evans+21, Evans+22}, with clouds being dominated by turbulence \citep[][]{ostriker+10, Evans+22}. In this scenario, which we refer to as { {\it turbulent support }(TS)}, MCs are in a quasi-equilibrium state between turbulence and self-gravity and, therefore, have long lifetimes ($\gg \tauff$) and a low stationary SFR \citep[e.g.,][]{MacLow_Klessen04, Krumholz_McKee05, Ballesteros-Paredes+07, McKee_Ostriker07, Krumholz_McKee20}. On the other hand, there is the global hierarchical collapse (GHC) scenario \citep{VS+09, Vazquez_Semadeni+19}, in which the gas in MCs is continuously flowing from the low-density towards the star-forming sites. In this model, the clouds evolve by growing in mass and size \citep{Heitsch_Hartmann08, Gonzalez-Samaniego+20, Vianey+20} by accretion from their surrounding material, increase their SFR \citep{Hartmann+12, ZA+12, ZA+14} and their massive-star content \citep{Hartmann+12, VS+17,VS+24}, and their lifetimes are determined by the time required to form stars sufficiently massive to interrupt the local SF episode and disperse the local gas.\\

In the { TS} scenario, it has been shown that, indeed, the parameter $\eff$ can be small ($\simeq0.01$) { and,} more importantly, { that it} depends on the virial parameter \cite[e.g.,][]{Padoan+12, Kim+21}.\footnote{ Note, however, that contrary to theoretical predictions, observations do not show a clear correlation between $\epsilon_{\rm ff}$ and $\alpha_{\rm vir}$ \citep[e.g.,][]{Leroy+25}.} Since a single cloud exhibits a variety of virial parameter values at different levels of its hierarchy \citep[at low surface densities, the virial parameter is typically large and approaches 1--2 at large surface densities, see Figs.~1 and 2 in][]{Evans+21}, it is hard to reconcile the constancy of $\eff$ found by \citet{Pokhrel+21} in the TS model.  \\

The SK law and the parameter dependence of $\eff$ in the context of collapsing clouds with stellar feedback have been { extensively} investigated recently. For example, both \citet{Kim+21} and \cite{suin2023stellar} studied the behaviour of $\eff$ for isolated, collapsing spherical clouds under the influence of feedback mechanisms. The former authors found that the $\eff$ depends inversely on the initial virial parameter, and both works found that the value of $\eff$ is strongly affected by the feedback and magnetic fields. Although we agree that stellar feedback is crucial in interrupting the star formation process, in the previous paper of this series, \citet{BP+23} have shown that the three results found by \citet{Pokhrel+21} (Eqs.~[\ref{eq:Pokhrel1}]-[\ref{eq:Pokhrel3}]), as well as the correlations shown in extragalactic studies \citep[e.g.,][]{Kennicutt98, Gao-Solomon04, Wu+05, Bigiel+08, Sun+23} can be understood simply as a consequence of the collapse of molecular clouds. In order to support this view, in the present contribution, we study the SK relation and the ratio of the SFR to the gas-infall rate ($\eff$) in numerical simulations of collapsing molecular clouds, following a methodology similar to that used for observations of local clouds. In \S \ref{sec:numerics} we describe the numerical simulations used in our analysis, which correspond to two filamentary structures. In \S \ref{sec:results} we present the results related to the SK relations and the parameter $\eff$. In \S \ref{sec:discussion} we discuss the implications of our results, while \S \ref{sec:conclusions} presents our main conclusions.

\begin{figure}
\includegraphics[width=0.5\textwidth]{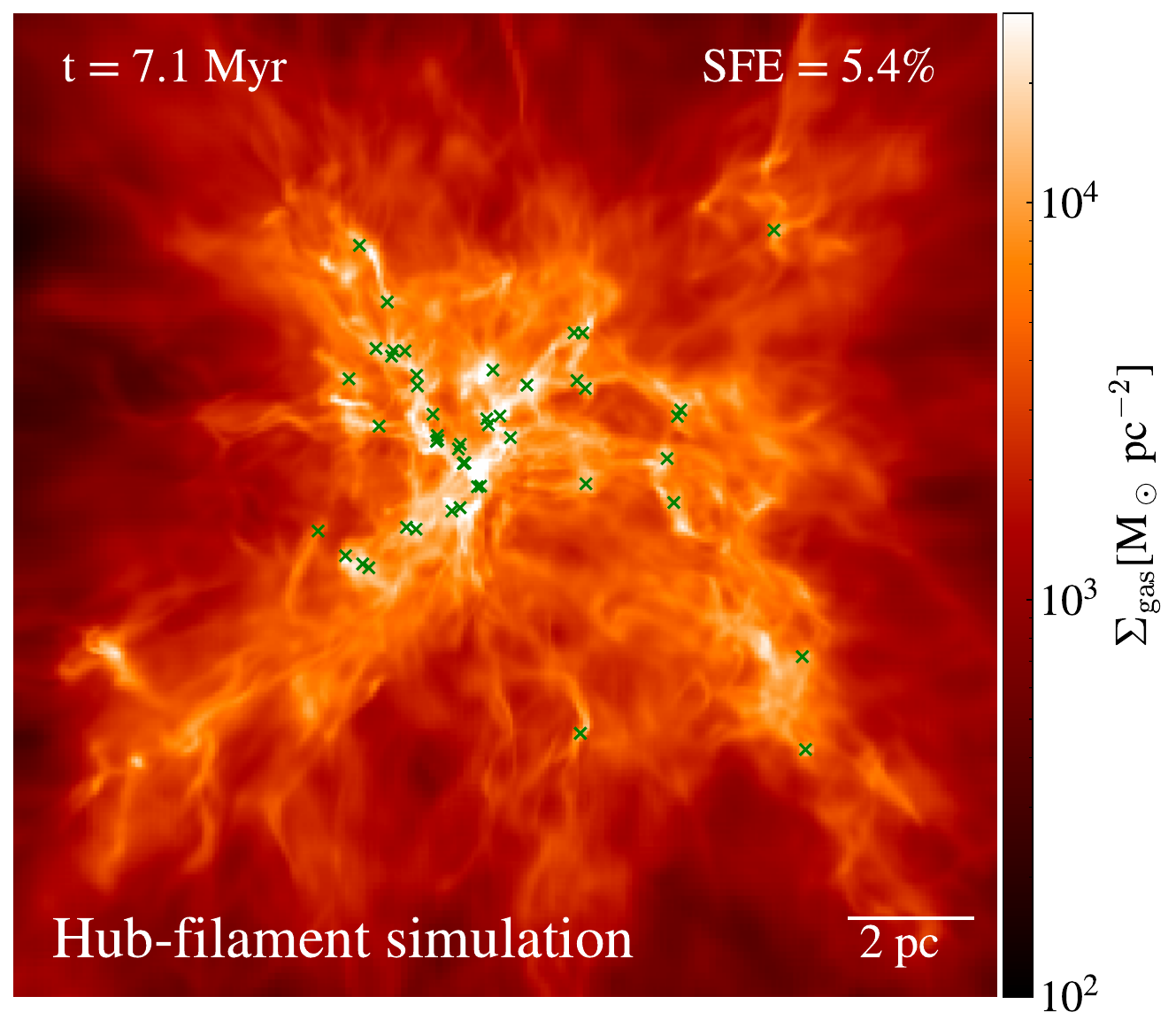}
\caption{Gas surface density of the ``Hub-filament'' simulation at $t=$7.1 Myr, %($\sim 0.94 \, \tauff$), 
time at wich the { star formation efficiency} is 5.4\% { (see the text)}. The green crosses (sink particles) represent groups of stars.}
\label{fig:filament-hub}
\end{figure}

\section{Numerical methods} \label{sec:numerics}

In this work, we study the SK relations in dense (molecular) structures dominated by self-gravity in the GHC scenario using three-dimensional magnetohydrodynamical (MHD) numerical simulations. \\

We use two different models of filamentary clouds undergoing collapse, which we refer to as the Filament simulation (FS) and the Hub-Filament simulation (HFS).  Both models include MHD, self-gravity, and the formation of sink particles. Feedback is not included. Both simulations have periodic boundary conditions either for the MHD, as well as for the gravity solver. These sets of simulations were performed with the AMR FLASH code \citep{FLASH1}. We use the MHD HLL3R solver \citep{Waagan+11} to solve the ideal MHD equations, ensuring a balance between accuracy and robustness, useful characteristics for highly supersonic astrophysical problems. To solve the Poisson equation for the self-gravity of the gas, we use an OctTree solver \citep{Wunsch+18}. We refine dynamically following the Jeans criterion \citep{Truelove+97}. A sink particle can be formed once a given cell reaches the maximum allowable refinement level and its density exceeds a specific density threshold ($\rho_{\rm thr}$). The formation of the sink particle involves multiple checks within a spherical control volume (of radius 2.5 times the size of the cell at the maximum level of refinement, $\Delta x$) around the cell of interest: 1) the gas is converging (negative divergence), 2) there is a minimum in the gravitational potential, 3) the cell is Jeans unstable, 4) it is gravitationally bound (negative total energy), and 5) the cell is not within the accretion radius of another sink \citep{Federrath+10}. The sink is created using the excess mass within the cell (that is, $M_{\rm sink} = (\rho - \rho_{\rm thr}) \Delta x^3$) and is capable of accreting mass from its surrounding environment. Note that given the size of the regions and the resolution of our simulations, a sink particle in both models represents a group of stars rather than an individual star. It is worth noting that the sink particle implementation in our simulations does not restrict the gaseous mass that can go into stars, as sink particles can accrete additional gas after formation, and new sinks can form until the collapsing gas is exhausted. This approach, based on the well-established scheme of \citet{Federrath+10} and widely used in star formation studies \citep[e.g.,][]{Padoan+12, Kim+19, Grudic+19}, accurately reproduces key features like the initial mass function.

Further details of the two simulations are as follows:\\

\begin{figure}
\includegraphics[width=0.5\textwidth]{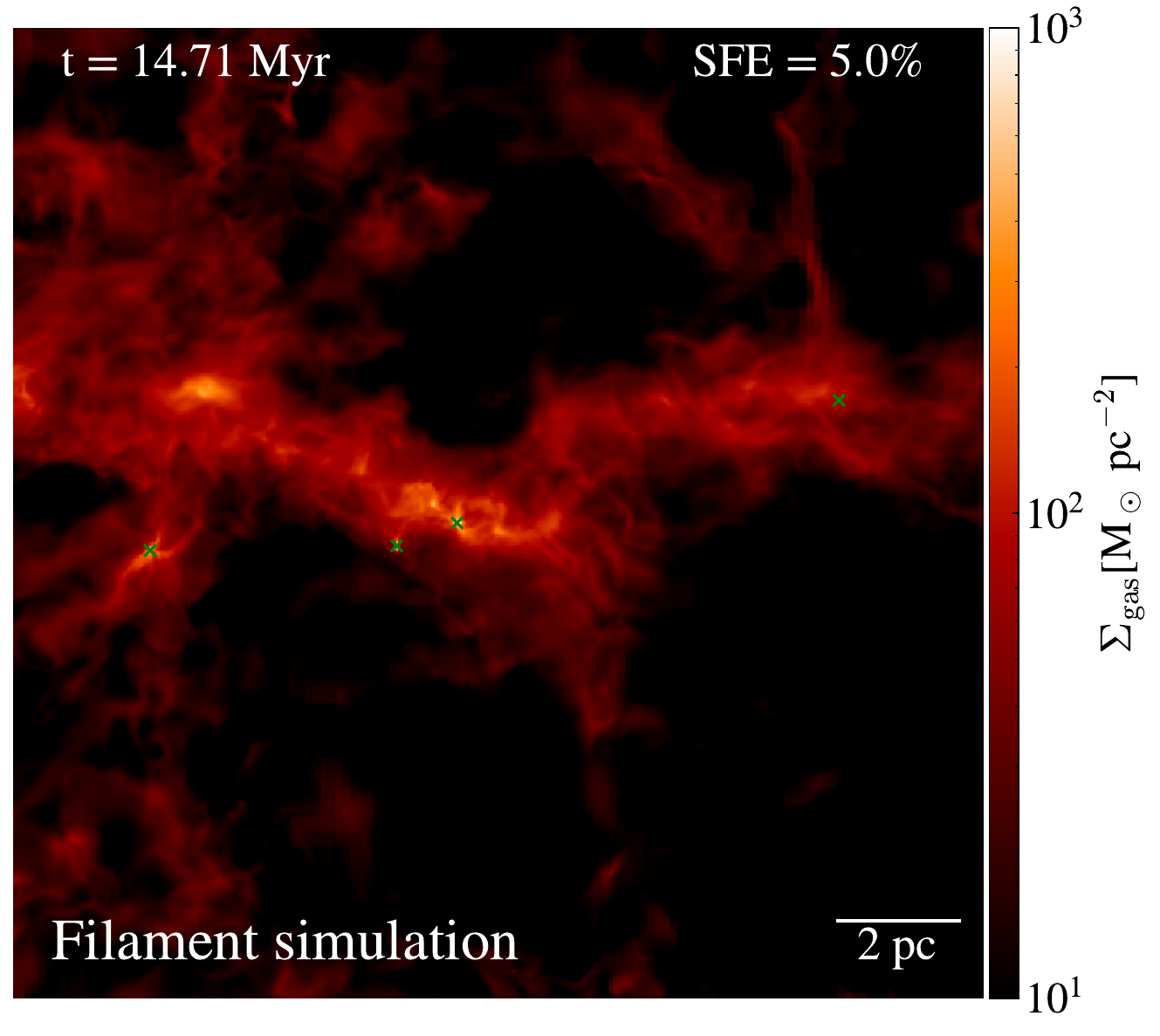}
\caption{Gas surface density of the ``Filament'' simulation at $t=14.7$ Myr, time at wich the star formation efficiency is 5\%. The green crosses (sink particles) represent groups of stars.}
\label{fig:filamento}
\end{figure}

{\bf Hub-Filament simulation (HFS)}. This simulation was first presented in \citet[][]{Carmen},\footnote{See also \citet{ZA+17}.} and follows the collapse of a cloud with an initial turbulent velocity field and constant density. Since the simulation is isothermal, we re-scale it to match the properties of local molecular clouds in the Solar Neighborhood. Thus, our simulation can be pictured as a hub-filament system with an initial linear size of 29.4~pc and an initial number density of $n_0 \sim 20$~cm\alamenos 3.\footnote{The corresponding mass density is $\rho = n \mu m_{\rm H} \sim 7.8 \times 10^{-23}$ g~cm$^{-3}$, with $m_{\rm H}$ the mass of a hydrogen atom and $\mu$ the mean molecular weight. For this model we assume $\mu = 2.3$, appropriate for molecular gas. This density implies a free-fall time of $\tau_{\rm ff,0}\simeq 7.59$~Myr.} This implies that, given a temperature of 10~K, the total mass is about $\sim$234 Jeans masses, with a total gass mass of $2.94\times 10^4 \, \Msun$. After $\sim$0.94~$\tau_{\rm ff,0}$, the linear size of its filaments converging to the central hub is about $\sim$~10~pc (see Fig.~\ref{fig:filament-hub}), comparable to the Monoceros molecular cloud \citep[see, e.g., ][]{Treviño-Morales+19}.

We initialize the simulation with a uniform magnetic field aligned along the $x-$axis, with a strenght of 1.7~$\mu$G, which makes the numerical domain magnetically supercritical\footnote{The mass-to-flux ratio is $\sim$6.7 times the critical value \citep[$\mu_{\rm crit} = (4 \pi^2 G)^{-1/2}$; e.g.,][]{Nakano_Nakamura1978}.} and prone to gravitational collapse. On the other hand, with over two hundred Jeans masses, the simulation is highly susceptible
to gravitational fragmentation. The initial turbulence in this simulation SEEDS (i.e., determines the location) of the sites of the local collapses, which occur at the same time as the collapse of the whole cloud. Since the initial geometry is a cube, it naturally forms filaments along the diagonals of the cube, with a density gradient towards the center, where a hub system is formed. \\

{\bf Filament simulation (FS)}. This run was presented in \citet[][model labelled as B3J]{ZA+18} to study the formation and evolution of magnetised dense structures. The simulation box, with sizes $L_{x}=256$, and $L_{y,z}=128$ pc, is filled with warm neutral gas at a number density\footnote{{For this model the initial mass density is $\rho_0 = n \mu m_{\rm H} \sim 4.2 \times 10^{-24}$ g~cm$^{-3}$, assuming $\mu = 1.27$, which corresponds to a helium fraction of 0.1 by number, appropriate for an atomic medium.}} of $2~\ppcc$ and a temperature of $1450~\K$. The initial magnetic field is uniform along the $x-$direction with a strength of 3~$\mu$G, so the formed cloud is magnetically supercritical. The thermodynamic behaviour of the ISM is modelled by incorporating heating and cooling processes, using analytic fits from \citet{Vazquez-Semadeni+07}, based on cooling functions from \citet{Koyama_Inutsuka02}. This results in a thermally unstable regime for densities 1--10~$\ppcc$ and temperatures 500--5000~K.\\

In this simulation, two cylindrical converging flows of radius $R=32$~pc and length $L=112$~pc are set to collide at a relative moderately supersonic velocity of 7.5~$\kms$. This collision forms denser, cold, neutral small clouds at the centre of the box, mainly due to the non-linear thermal instability triggered by the compression. The newborn cloud continues to accrete gas from the cylinders and eventually becomes dense enough to be considered molecular. The resulting morphology of the newborn molecular cloud is characterised by a network of filaments, and in this work, we focus on one particular filament previously analysed by \citet[][see Fig.~\ref{fig:filamento}]{Vianey+23}. This filament is contained in a sub-box of 16 pc in size. The mass in dense gas increases from $\sim 3000 \Msun$ to $\sim 4000 \Msun$ during the period of our analysis.

In our analysis of both simulations, we define the zero time point as the moment the first sink particle forms, occurring 5.6 Myr and 11.2 Myr after the simulation start for the HFS and FS, respectively. At this zero times, we calculate the free-fall times of 2.97 and 2.87~Myr using the mean density of gas that would be considered molecular (i.e., gas with density $n>$100~cm\alamenos3).\footnote{Since our study focuses on the molecular phase and we do not include chemical processes in our simulations, we consider gas with densities above this threshold ($100 \pcc$) as molecular, as it represents the typical average density of MCs reported in the literature \citep[see e.g.,][]{Murray2011}.} This corresponds to mean densities of $\promedio{n (t=5.6 \, {\rm Myr})}=$~139.8 and $\promedio{n (t=11.2 \, {\rm Myr})}=240.9 \,\ppcc$, for the HFS and FS, respectively.

In both simulations, as expected from theory \citep[e.g., ][]{Franco_Cox86, Hartmann+01}, while the compressed gas becomes denser and more massive, the clouds become gravitationally unstable and proceed to collapse. { A comparison of Figs. \ref{fig:filament-hub} and \ref{fig:filamento} shows that the HFS has a higher average density compared to the FS, making it more strongly gravitationally bound.} And just as it occurs with a system of particles \citep[e.g.,][]{Noriega-Mendoza_Aguilar18}, the natural outcome of the gravitational collapse of a molecular cloud is to exhibit equipartition between its gravitational and internal energies \citep[see also][]{Vazquez-Semadeni+07, Noriega-Mendoza_Aguilar18, Vianey+23, Ibañez-Mejia+22}, which may very well be mistaken as virial equilibrium \citep{Ballesteros-Paredes+11a, Ballesteros-Paredes+18, Vianey+23}.\footnote{ Note that while the condition $E_k \sim -E_g$ is often interpreted as virial {\it equilibrium}, this is not necessarily true. Virial equilibrium is defined as the condition $\ddot{I} = 0$ in the virial theorem, where $I$ is the moment of inertia of the system. However, the virial theorem is intrinsically a time-dependent equation, and the condition $E_{\rm k} \sim -E_{\rm g}$ can also arise in dynamical, nonequilibrium situations (i.e., in which $\ddot I \ne 0$). In dissipationless $N$-body systems, such as globular clusters, the collapse can be stopped by virialization because the kinetic energy generated by the collapse can be stored in the system itself. Indeed, \cite{Noriega-Mendoza_Aguilar18} showed that during the gravitational collapse of initially subvirial systems, $E_k \sim -E_g$ is achieved near the free-fall timescale due to the dissipationless nature of particle systems. On the other hand, in dissipative and radiative gaseous systems, the energy cannot be stored, and the collapse continues, implying that $\ddot{I} \ne 0$ . Nevertheless, in such systems, the kinetic energy in the {\it infall} motions can still satisfy $E_{\rm k} \sim E_{\rm g}$ \citep[e.g.,] [] {Vazquez-Semadeni+07, Guerrero_Gamboa+20}, because the kinetic energy is extracted from the gravitational energy.}

\subsection{Structure selection and computed quantities} \label{subsec:structure}

Since we want to mimic the observational procedure to estimate dense gas mass surface densities, we projected the volumetric density along one of the axes. This gives us the mass surface density on the plane defined by the other two axes. We then identify iso-surface density contours and measure the mass in dense gas ($\Mmol$), the mass in newborn stars ($\Msinks$), and the area ($A$) inside them. From these quantities, we can compute the size of the cloud ($R$) and the mass density ($\rho$) as
\begin{equation} \label{eq:r}
    R = \sqrt{\frac{A}{\pi}},
\end{equation}
\begin{equation} \label{eq:rho}
  \rho = \frac{3 \pi^{1/2}}{4}\ \frac{\Mmol}{A^{3/2}}.
\end{equation}
With this procedure, we can compute different quantities related to SK-type relations:\\

\begin{enumerate}

\item { The { instantaneous} star formation efficiency (SFE)}. At all times, we estimate the total {instantaneous} SFE {in the raw 3D simulations} as 
\begin{equation} \label{eq:sfe}
  \SFE(t) = \frac{\Msinks(t)}{\Mmol(t) + \Msinks(t)} ,
\end{equation}
with $\Msinks(t)$ being the total mass in sink particles at time $t$ within the projected structure and $\Mmol(t)$ being the dense gas mass ($n>100 \, \ppcc$).  \\

 Furthermore, as pointed out earlier, our simulations do not include stellar feedback.\footnote{{ Note that pre-supernova feedback mechanisms such as jets, winds, ionizing radiation, or radiation pressure, could offer additional support against gravitational collapse and influence the overall dynamics of MCs, eventually leading to the disruption of the cloud.}} However, to avoid unrealistic results due to its absence, we analyse the simulations at early stages, when the SFE is 2\%, 5\%, and 10\% (i.e., at $t=$1.2, 1.5, and 1.7 Myr) for the HFS. For the FS, we analyse times when the SFE is 5\%\ and 10\% (i.e., 3.5 and 4.3 Myr).\footnote{For the FS, we do not analyse the time at which the SFE is 2\%\ since it only has two sinks.}

\item
{ The star formation rate}. We note that in observational works, the SFR is typically estimated as the mass that has gone into some type of proto-stellar object, $M_*$, divided by the typical life span of this type of objects, $\tausf$ \citep[e.g.,][]{Lada+10, Pokhrel+21}; that is
\begin{equation}
    \SFR \approx \frac{M_*(t)}{\tausf}.
    \label{eq:sfr}
\end{equation}

Observationally, the mass in newborn stars is usually computed as the number of protostellar objects multiplied by the typical mass of the protostar, $\promedio{M}$~$\sim$~0.5$~\Msun$. In the simulations, we have the advantage of knowing exactly the amount of mass that has gone into sink particles, $\Msinks$, so we use this quantity directly, i.e., $M_*$~$=$~$\Msinks$ at every timestep. \\

As for the timescale, characteristic timescales have been estimated for protostellar objects at different stages \cite[e.g.,][]{Ward-Thompson+07, Heiderman-Evans-2015}. If the objects to count are generic YSOs, the assumed characteristic timescale is $\tausf$~$\sim$~2~Myr \citep[e.g.,][]{Evans+09, Heiderman+10, Lada+10}. In contrast, when considering embedded protostellar objects, the characteristic timescale is of the order of $\tausf$~$\sim$~0.5~Myr \citep[e.g.,][]{Pokhrel+21, Lada+13}. \cite{Dib+2024} suggest that using less-evolved embedded objects (Class 0) gives a more reliable estimation of the SFR, especially for a time-dependent burst-like star formation history. 

In order to mimic observations, in our simulations we want to compute the mass accreted by sink particles since they first appear. Consequently, we compute $\tausf$ as the time spent between the formation of the first sink particle and the time under analysis. In the HFS, $\tausf$ is 1.2, 1.5, and 1.7 Myr for SFEs of 2\%, 5\%, and 10\%, respectively. In the FS, $\tausf$ reaches 3.4 and 4.2 Myr at SFEs of 5\% and 10\%, respectively.\\

\item
{ The ratio of the SFR to the gas-infall rate ($\eff$)}. This parameter can be computed as the ratio between the averaged $\SFR$ (Eq. \ref{eq:sfr}) and the {\rm free-fall} collapse rate, ($\Mmol/\tff$), with $\Mmol = \Sigmamol*A$ \citep[e.g.,][]{Ballesteros-Paredes+23}. 

\end{enumerate}

\begin{figure*}
\includegraphics[width=0.45\textwidth]{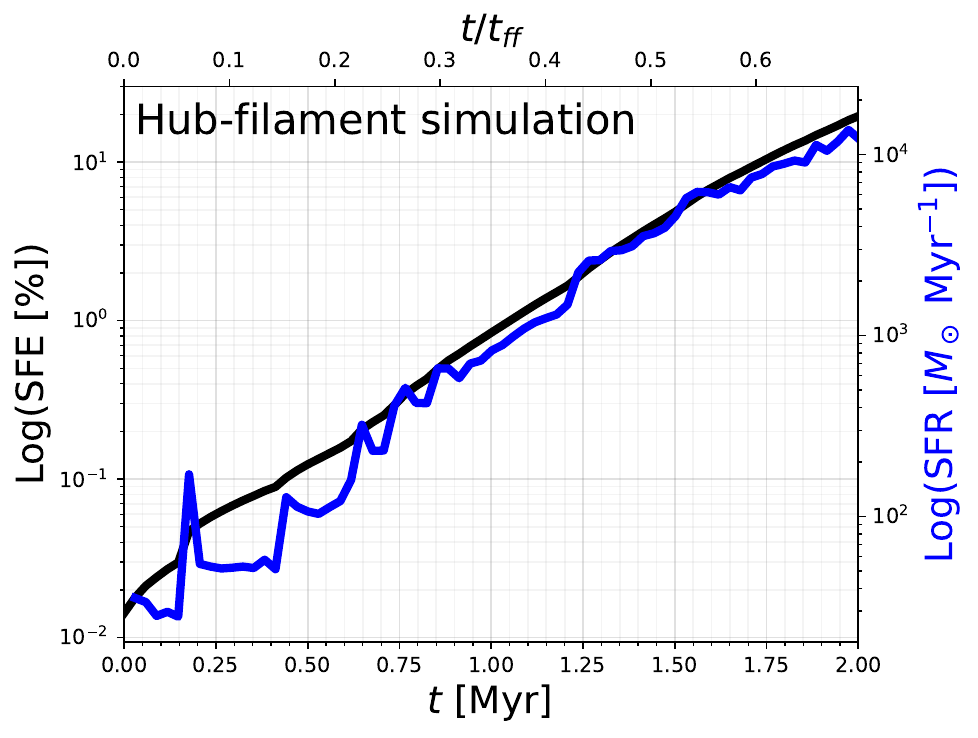}
\includegraphics[width=0.45\textwidth]{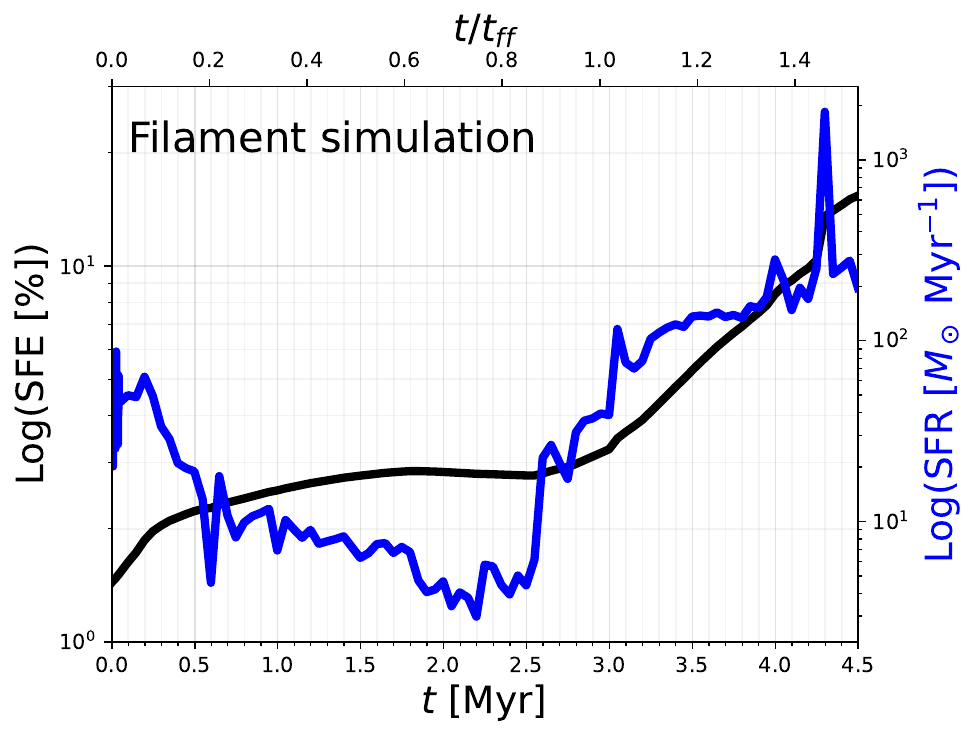}
\caption{Star formation efficiencies (black lines, see eq.~\ref{eq:sfe}) and star formation rates (blue lines, see eq.~\ref{eq:sfr}) as a function of time for the Hub-filament simulation (left) and the Filament simulation (right). The top axes show the time in units of the free-fall time. We set the zero time at the onset of sink formation and calculate the free-fall time using the mean density of dense gas (with $n>100 \, \ppcc$) at this time (see \S \ref{subsec:SFE}).
}
\label{fig:efficiencies}
\end{figure*}

\section{Results}\label{sec:results}

Our ultimate goal is to understand the origin of the Schmidt-Kennicutt relation, the constancy of $\eff$, and its observed low intracloud value ($\eff \sim 3 \times 10^{-2}$) { in resolved clouds as reported by \citet{Pokhrel+21}}. For this purpose, in this section we analyze, for each one of our simulations, the evolution of the instantaneous star formation efficiency, the SK relations, and the ratio of the star formation rate to the gas-infall rate, $\eff$.\footnote{We avoid to use the term ``efficiency per free-fall time'' because as we explain in \S\ref{sec:intro}, it is clear it is not an efficiency.}

\subsection{The instantaneous Star Formation Rate and  Star Formation Efficiency} \label{subsec:SFE}

Fig.~\ref{fig:efficiencies} shows the time history of the SFE (eq. \ref{eq:sfe}) for the HFS (left panel) and the FS (right panel) simulations, measured directly from the 3D raw simulation data. The upper axes of this figure show the time in units of the corresponding free-fall times (see \S\ref{sec:numerics}). 

As can be seen, in both cases, the { instantaneous} SFE increases over time, reaching large values within one free-fall time of the initial gas density. This behaviour is due to the absence of stellar feedback in both simulations or the limited supply of gas within the simulation domain available for the cloud to accrete. Although both curves share some similarities, they exhibit an important difference, the HFS (panel a) simulation reaches 20\%\ efficiency within a fraction of the free-fall time. In contrast, the FS (panel b) takes twice that time (in units of the free-fall time) to reach a similar value. On the other hand, note that the final 2 Myr of evolution of the FS simulation behave very similarly as the entire duration of the HFS one, suggesting that, after this time, the FS simulation has engaged in a similar regime as HFS.

Fig.~\ref{fig:efficiencies} also shows the SFR over time (blue solid lines, right $y$-axes). Similarly to the SFE, the SFR exhibits substantial growth in both cases, although in the FS case (see Sect. \ref{sec:low_eff}), the growth is modulated by accretion flows into the cloud from larger scales, which inject mass at low density into the clouds, without an instantaneous increase of star formation, which takes longer to occur. The substantial growth of the SFE is the product of gravitational collapse \citep[e.g.,][]{Zamora-Aviles+18}, and it is { expected to continue} until feedback from massive stars destroys and disperses the cloud \citep[see, e.g.,][]{ZA+12, Colin+2013, suin2023stellar, Guszejnov+2022}. \\

It is important to note that the masses of real molecular clouds hardly remain constant as they evolve. As numerical work has shown since long ago, molecular clouds are in a continuous dynamical exchange of mass, momentum, and energy with their surroundings \citep[e.g., ][]{Bania_Lyon80, Vazquez-Semadeni+95, Ballesteros-Paredes+99a}. In particular, molecular clouds are likely to be formed by the convergence of large-scale diffuse flows \citep{Ballesteros-Paredes+99a, Hennebelle_Perault99, Heitsch+05, VS+06}, and as a natural consequence, their masses evolve continuously \citep[e.g.,][]{Heitsch_Hartmann08, VS+09, ZA+12,ZA+14}. In the FS, the mass of gas with number density grater that $100 \, \ppcc$ is $\sim 2.4 \times 10^3 \, \Msun$  at the onset of star formation and increases to $\sim 3 \times 10^3 \, \Msun$ when the SFE reaches 10\%. Indeed, in both of our simulations, the dense (with $n > 100 \, \ppcc$) gas mass increases over time, as it accretes from its lower-density surroundings. This has the important consequence that, although the stellar mass increases over time, so does the dense gas mass, and therefore the instantaneous SFE does not increase as much \citep[see also][]{Gonzalez-Samaniego+20}. \\

%%%%%%%%%%%%%%%%%%%%%%%%%%%%%%%%%%%%%%%%%%%%%%%%%%%%%%%%%%%

\begin{figure*}
  \includegraphics[width=\textwidth]{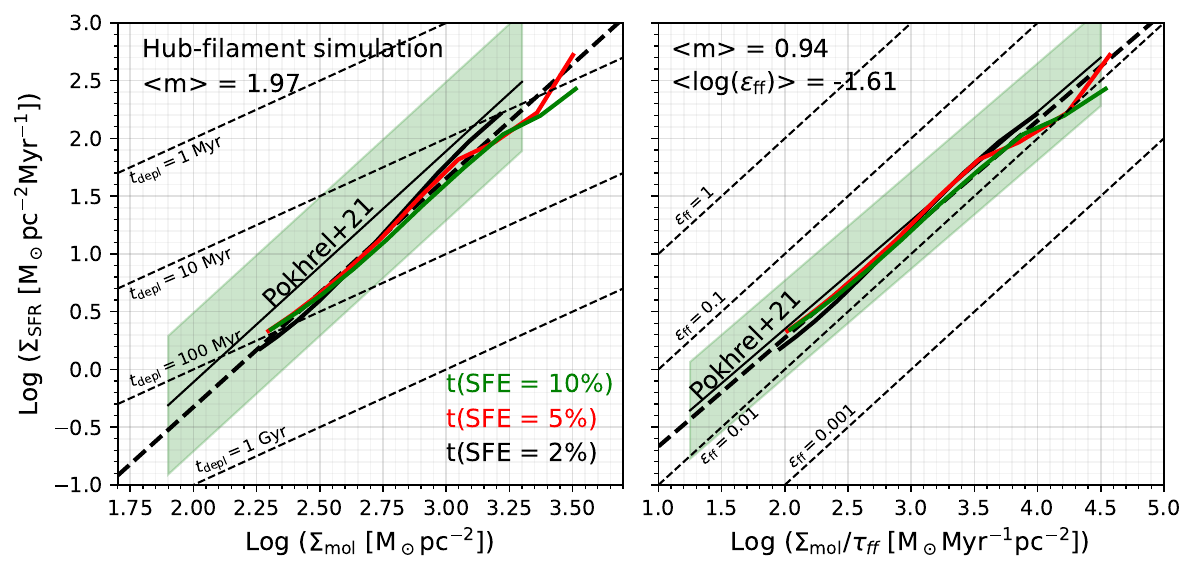}
  \caption{The SK relation for the simulated Hub-filament system, quantified at three different times, when the SFE reaches 2\%, 5\%, and 10\% (black, red, and green lines). The left panel shows the relationship $\Sigmasfr - \Sigmamol$, while the right panel shows the relationship and $\Sigmasfr - \Sigmamol/\tauff$. { Each data point along a line} corresponds to a different surface density isocontour of the same structure, as in the observational data. We also compute the corresponding linear fits for each curve and show the average slope of the three different lines, $\average{m}$. In addition, the right panel shows the mean value of the three intercepts, $\average{\log{\eff}}$.
  The thin dashed lines in the left panel represent, from top to bottom, depletion times $\taudepl$ of 1, 10, 100, and 1000 Myr. Similarly, in the right panel, thin dashed lines represent $\eff$ of 1, 0.1, 0.01, and 0.001 (from top to bottom). In both panels, the thin solid lines represent the fit reported in \citep{Pokhrel+21}, while the green shadings indicate their reported spread. 
    }
   \label{fig:Core-KS}
\end{figure*}

\subsection{SK type relations of collapsing clouds}

\begin{figure*}
	\includegraphics[width=\textwidth]{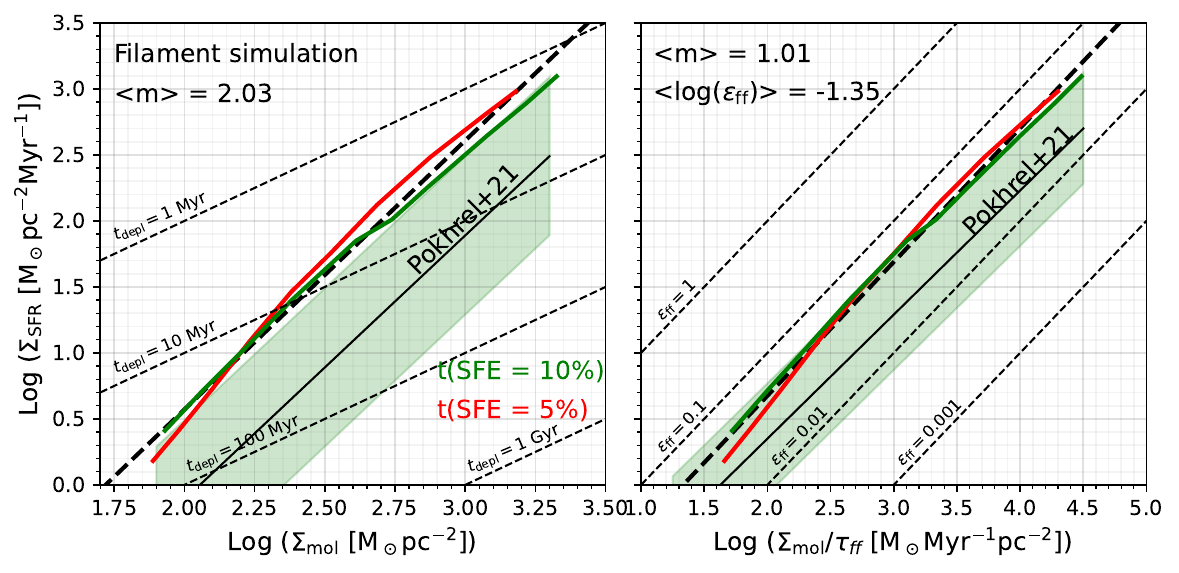}
    \caption{Same as Fig. \ref{fig:Core-KS}, but for the simulated Filament. 
    }
   \label{fig:Fil-KS}
\end{figure*}

Our main goal in this contribution is to locate our clouds in the $\Sigmasfr-\Sigmamol$ and $\Sigmasfr-\Sigmamol/\tauff$ diagrams. Thus, we need to compute the area, SFR, gas mass, and free-fall time at each closed contour as we described in \S\ref{subsec:structure}. As explained before, we compute $\tausf$ as the time since the formation of the first sink to the time when the analysis is performed, corresponding to total efficiencies of 2\%, 5\%, and 10\%. The results are shown in Figs.~\ref{fig:Core-KS} for the HFS and \ref{fig:Fil-KS} for the FS. In these figures, we additionally show lines that denote constant depletion time (thin dashed lines with slope equal to one, left panels) and lines denoting constant 
ratio of SFR to gas-infall rate, $\eff$ (thin dashed lines with slope equal one, right panels).  From these figures, we find the following scalings:\footnote{We use a linear least squares fit in the log-log space.}\\

\begin{enumerate}
  
  \item  $\Sigmasfr = a \Sigmamol^{N_1}$, where $a$ and $N_1$ are the mean intercept and slope, respectively. $a = 5.2 \times 10^{-5}$ and $N_1 =1.97$ for the HFS. Similarly, $a=3.1 \times 10^{-4}$ and $N_1=2.03$ for the FS. \\
    
  \item $\Sigmasfr = \average{\eff}$ ($\Sigmamol/\tauff)^{N_2}$, with $\average{\eff}$ and $N_2$ the mean intercept and slope, respectively. $\average{\eff}=0.030$ and $N_2=0.94$ for the HFS. Similarly, $\average{\eff}=0.045$ and $N_2=1.01$ for the FS.
  
\end{enumerate}

These values are consistent with the slopes reported for resolved clouds \citep[e.g.,][]{Pokhrel+21} and the $\Sigmasfr - \Sigmamol/\tauff$ relation is tighter than the $\Sigmasfr - \Sigmamol$ relation, as it occurs in observations \citep{Pokhrel+21}. \\

A third point to notice from the left panels of Figs.~\ref{fig:Core-KS} and \ref{fig:Fil-KS} is that, the denser the cloud, the { shorter} the depletion times (left panels), although $\eff$ remains constant (right panels). We discuss this point further in \S \ref{sec:low-eff}.

%%%%%%%%%%%%%%%%%%%%%%%%%%%%%%%%%%%%%%%%%%%%%%%%%%%%%%%%%%%%%%
\subsection{The ratio of the SFR to the gas-infall rate ($\eff$)}

\begin{figure}
\includegraphics[width=\columnwidth]{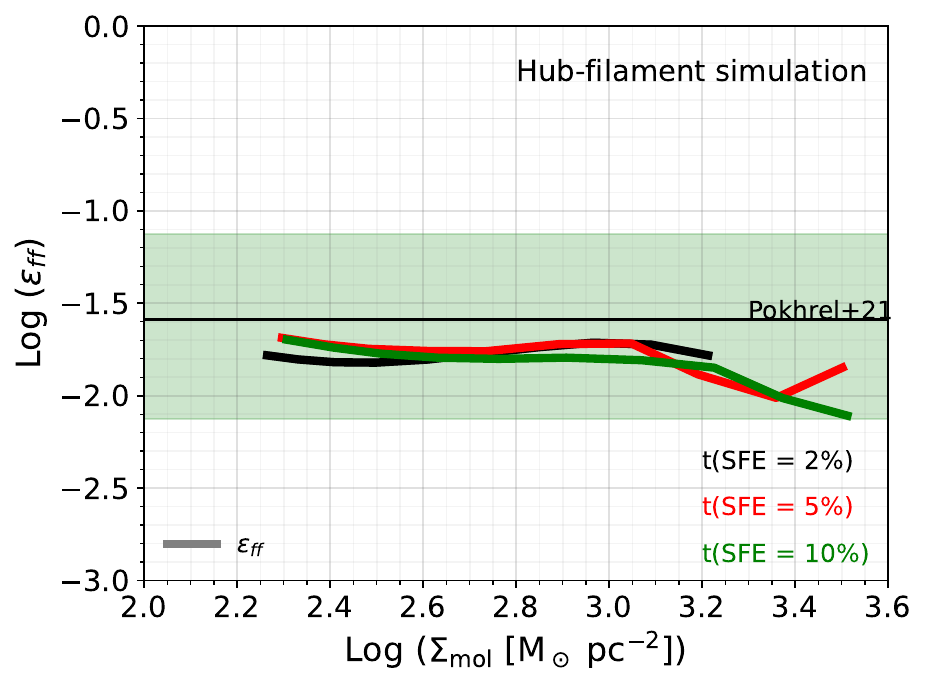}
\caption{The $\eff$ as a function of surface density for the Hub-filament system  measured at three different times (SFE of 2\%, 5\%, and 10\%, represented by black, red, and green continuous thick lines, respectively). The horizontal line represents the average observational determination by \citet{Pokhrel+21}. The green band represents the range of values reported by these authors { including their uncertainties.}}
\label{fig:Core-eff}
\end{figure}

\begin{figure}
\includegraphics[width=\columnwidth]{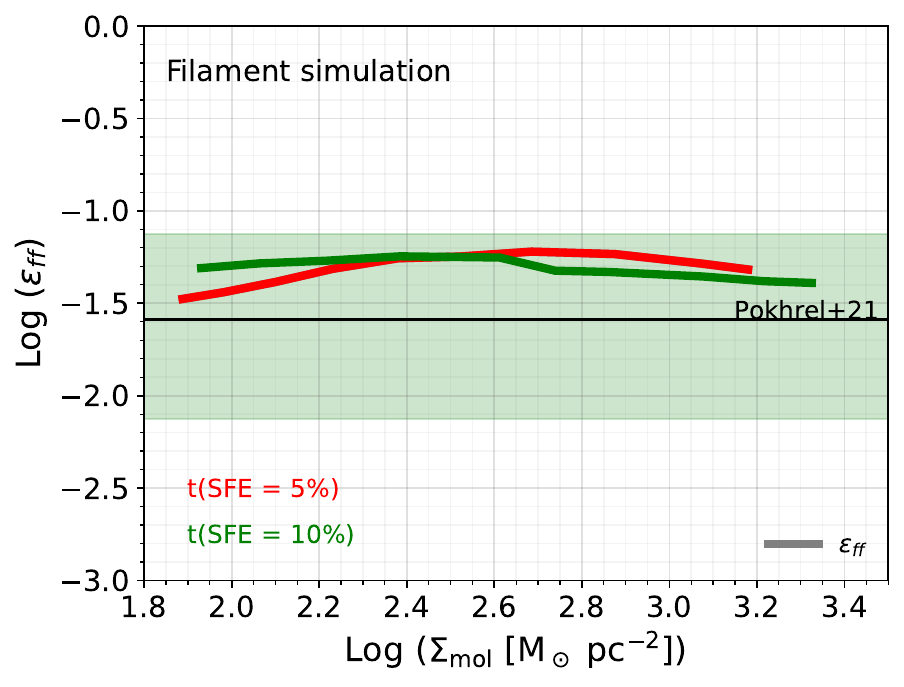}
\caption{Same as Fig. \ref{fig:Core-eff} but for the Filament simulation.} 
\label{fig:Fil-eff}
\end{figure}

In addition to the SK-type relations, we also compute $\eff$ (Eq. \ref{eq:eff}) as described in \S\ref{subsec:structure}, for different surface density thresholds within each cloud. The results are shown in Figs. \ref{fig:Core-eff} and \ref{fig:Fil-eff}, for the HFS and the FS, respectively.\\

Several points are noteworthy in these figures. First, we note that the HFS simulation (Fig.~\ref{fig:Core-eff}) has a significantly smaller value of $\eff$ ($\sim 2\%$) than the FS ($\eff \sim 5\%$, Fig.~\ref{fig:Fil-eff}), in spite of  having a significantly larger instantaneous SFE at a given fraction of the free-fall time ($\sim 10\%$ versus $\sim 3\%$ at $t \approx 0.6 \tauff$). This suggests that $\eff$ is not an accurate measure of the actual SFE. Second, we note that $\eff$ is nearly independent of the surface density threshold, as could be inferred from Figs. ~\ref{fig:Core-eff} and \ref{fig:Fil-eff}. This behaviour is consistent with the fundamental law of star formation (eq. \ref{eq:fundamental}). In addition, the range of values computed for $\eff\in(0.014, 0.063)$ in both simulations is consistent with the observational counterpart \citep[][green shaded region in Figs.~\ref{fig:Core-eff} and \ref{fig:Fil-eff}]{Pokhrel+21}. In addition, we also note that $\eff$ remains relatively constant over time in both simulations, even though the instantaneous efficiency is increasing from 2\% to 5\% and 10\%. Our results are consistent with those of \citet{Pokhrel+21}, who found no systematic difference in $\eff$ between clouds with different star formation activity. In Sections~\ref{sec:low_eff} and \ref{sec:const_eff}, we explain this behaviour as a direct consequence of gravitational collapse.

\section{Discussion}\label{sec:discussion}

\subsection{Implications of an increasing SFR and a collapse-determined density profile on the measurement and meaning of $\eff$} 
\label{sec:low-eff}

\subsubsection{The effect of accretion on the estimated structure lifetimes} \label{sec:effect_of_accr} 

For over nearly 50 years, star formation has been known to be inefficient \citep[e.g., ][]{Zuckerman_Evans74, Krumholz_Tan07, Evans+22} because, on galactic levels, it may take $\sim$2~Gyr to convert all molecular gas currently present in the Milky Way into stars, at the current star formation rate \citep[e.g., ][]{Bigiel+08, Kennicutt_Evans12}, while a simple estimate of the ``free-fall SFR'', allowing all molecular gas in the Galaxy to form stars in one free-fall time, would be $\sim 100$ times larger than the observed one. As a consequence, some works have considered star formation to be a ``slow'' process \citep[e.g., ][]{Krumholz_Tan07, Evans+21, Evans+22}. This argument is at the core of the turbulent picture of molecular cloud dynamical evolution, in which molecular clouds, having large Jeans numbers, have low star formation efficiencies because they are assumed to be supported against collapse by supersonic turbulent motions over several free-fall times.\\

{ In the context of the model of MCs supported by turbulence against collapse (TS),} \citet{Krumholz_McKee05} introduced the parameter $\eff$ (eq.~\ref{eq:eff}) as a measure of the star formation efficiency over one free-fall time, picturing clouds as entities that live many free-fall times supported against collapse by turbulence and magnetic fields and have a constant SFR. In this way, { we can estimate, as a first approximation, that} if a cloud had a { constant} efficiency per free-fall time of $\eff\sim0.01$, the present efficiency after the assumed time span of a cloud \citep[$\sim$10 free-fall times, see e.g.,][]{Krumholz+19}, would be SFE~$\sim$~0.1, compatible with observational estimations of the star formation efficiency \citep[e.g.,][]{Myers+86}. { That is, under the TS model,} the low measured values of $\eff$ have been taken as evidence that clouds not only are long-lived\footnote{{ By "long-lived," we refer to a duration spanning several free-fall timescales.}} and supported against collapse by turbulence, but also that star formation is governed and regulated by local processes, such as turbulence or stellar feedback \citep[e.g., ][]{Pokhrel+21, Millstone+23}. \\

However, the GHC model \citep{Hartmann+01, Vazquez-Semadeni+07, VS+09, Vazquez_Semadeni+19, Heitsch_Hartmann08, Ballesteros-Paredes+11a, Ballesteros-Paredes+18, Ibañez-Mejia+17, Ibañez-Mejia+22} proposes the alternative point of view that the gas within MCs is not systematically supported against collapse but rather is continuously {\it flowing} from the low-density regions (cold HI clouds and CO-dark molecular gas) toward the high-density ones (clumps, hub-filament systems, cores, and ultimately, stars). In this context, MCs are "segments" (stationary parts) of a continuous accretion flow from low to high density. Moreover, numerical simulations indicate that, since observationally, the clouds are defined by column density thresholds, they have an intense exchange of mass, momentum, and energy with their surrounding \citep{Ballesteros-Paredes+99a}. In particular,  masses increase due to the accretion from their environment \citep[][see also the analytical model of  \citealt{ZA+12}] {VS+09, Gonzalez-Samaniego+20, Vianey+20}. As the clouds' masses increase, so do their SFR and the masses of the most massive stars they contain \citep{ZA+14, VS+17, Vianey+20, VS+24}, until massive stars capable of destroying the clouds appear several Myrs after the onset of star formation.\\

Within this framework, the clouds and cores appear to have long lifetimes not because they are somehow supported against collapse by some agent but rather because they continue to accrete material while they are forming stars, so that they are continuously being replenished, until the time at which they are disrupted. Indeed, there is plenty of evidence that after ~$\sim 5$~Myrs, a timescale that is $\sim 1.5$ times the free-fall time at densities $\sim$100~cm\alamenos3, the characteristic density of CO molecular clouds, stars disperse their parental cloud \citep[e.g., ][]{Herbig78, Leisawitz+89, Ballesteros-Paredes+99b, Hartmann+01, Ballesteros-Paredes_Hartmann07}. This timescale is in agreement with more recent estimations of cloud lifetimes in extragalactic studies \citep{Kruijssen+18}.

\subsubsection{Problems with the interpretation of $\eff$ as an efficiency in accreting and evolving clouds} \label{sec:problems_eff}

In the aforementioned dynamical and evolutionary context of the GHC, $\eff$ loses significance as an ``efficiency'', where by efficiency we mean the fraction of the available gas mass that is transformed into stars during one free fall time of the cloud, because of mainly two reasons:

\begin{enumerate}
    \item The molecular gas reservoir is not fixed but rather increases over time as the cloud accretes from its environment \citep[e.g., ][]{Kawamura+09, Lee+16}. Thus, the notion of the ``available gas mass'' is undefined, as the gas mass is continuously replenished. \\

    \item The rate at which stars form in the clouds and clumps (the SFR) is not constant, but instead increases over time until the time when feedback from the newly-formed stars begins to erode the structures, at which point the SFR begins to decrease. Therefore, the rate at which the gas is consumed to form stars is not constant but rather varies in time. \\

\end{enumerate}

In addition, it is important to { note} that it is not hard to find configurations where $\eff>1$ \citep[see, e.g., Fig.~7 in ][]{Clark_Glover14}. But by definition, an efficiency must be smaller than unity. Indeed, assuming that $\eff$ were an efficiency, $\eff>1$ would mean that the mass converted into stars within one free-fall time would be larger than the mass available for star formation, violating mass conservation. Thus, it is best to interpret $\eff$ as a ratio of two {\it instantaneous} rates, as indicated by eq.\ \eqref{eq:eff}.

\begin{figure*}
\includegraphics[width=0.45\textwidth]{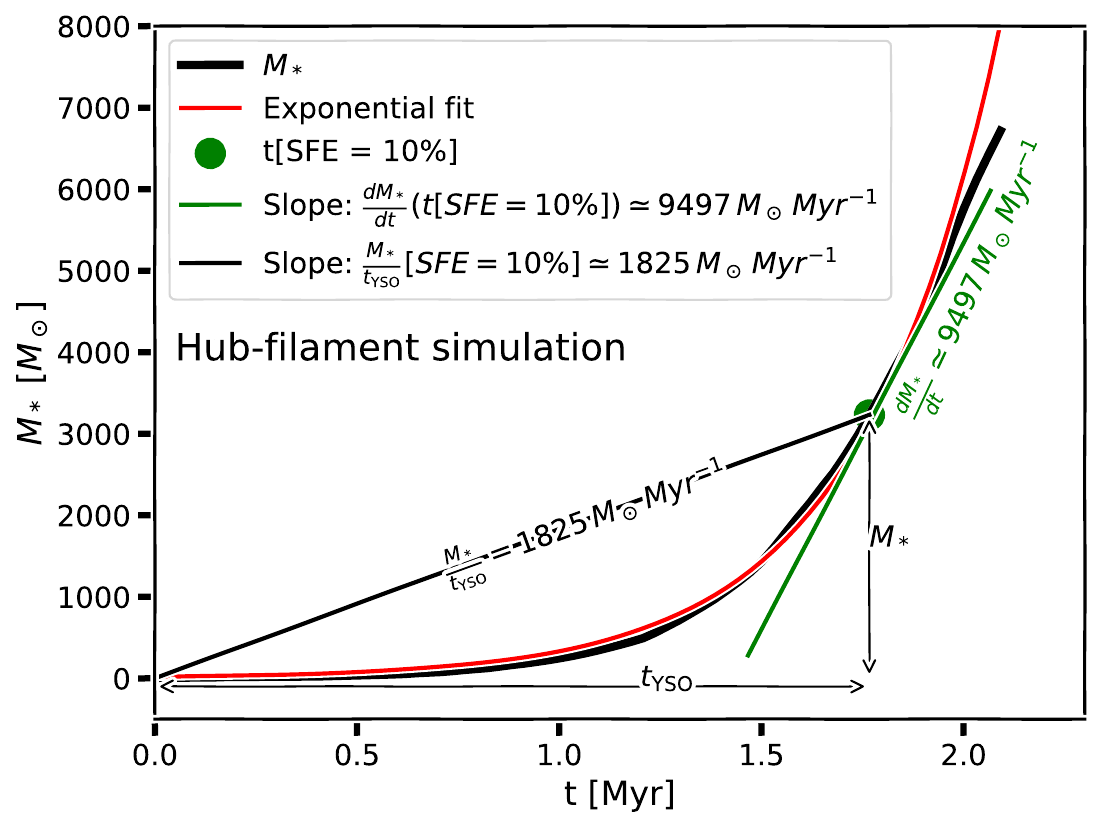}
\includegraphics[width=0.45\textwidth]{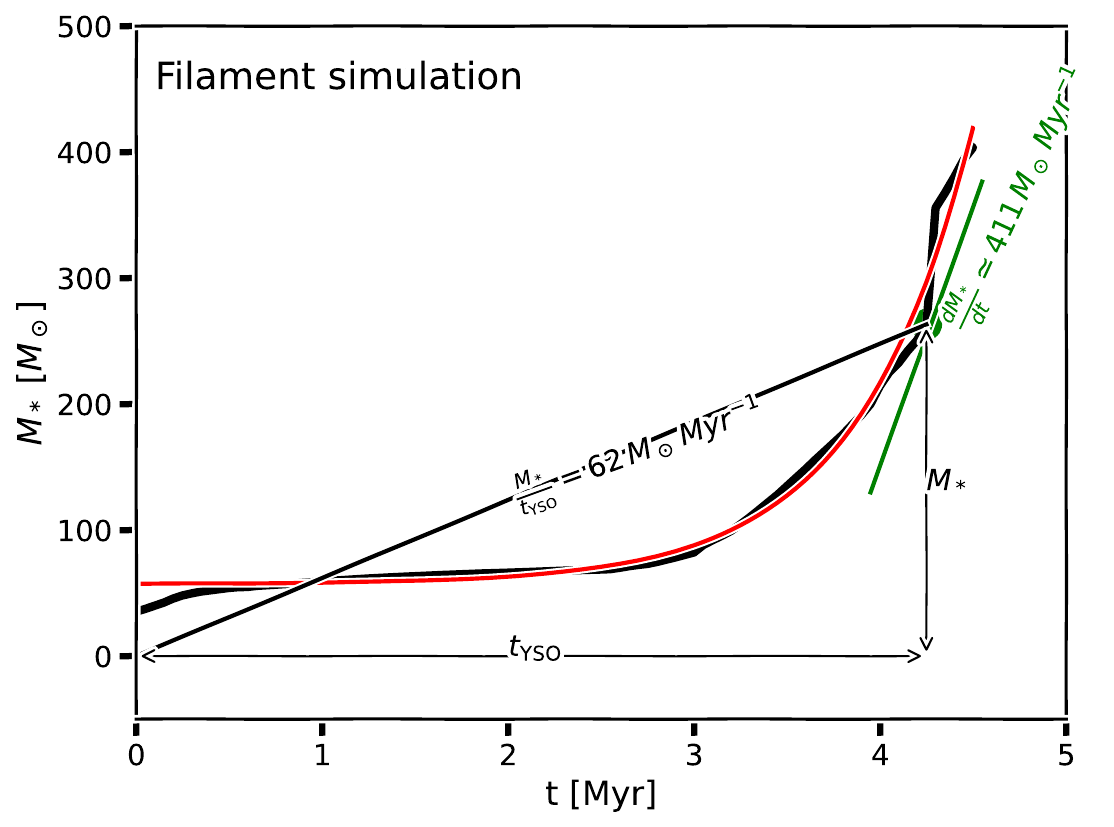}
\caption{
Mass of sinks as a function of time for the HFS (left panel) and the FS (right panel). The thick black lines represent direct measurements from the simulations. The red lines indicate exponential fits to the data, while the green lines show the slope at the point where the star formation efficiency (SFE) reaches 10\%. Note that the average SFR (represented as $M_* / \tausf$) is necessarily lower than the instantaneous SFR ($dM_*/dt$).
}
\label{fig:Msinks}
\end{figure*}

\begin{figure}
\includegraphics[width=\columnwidth]{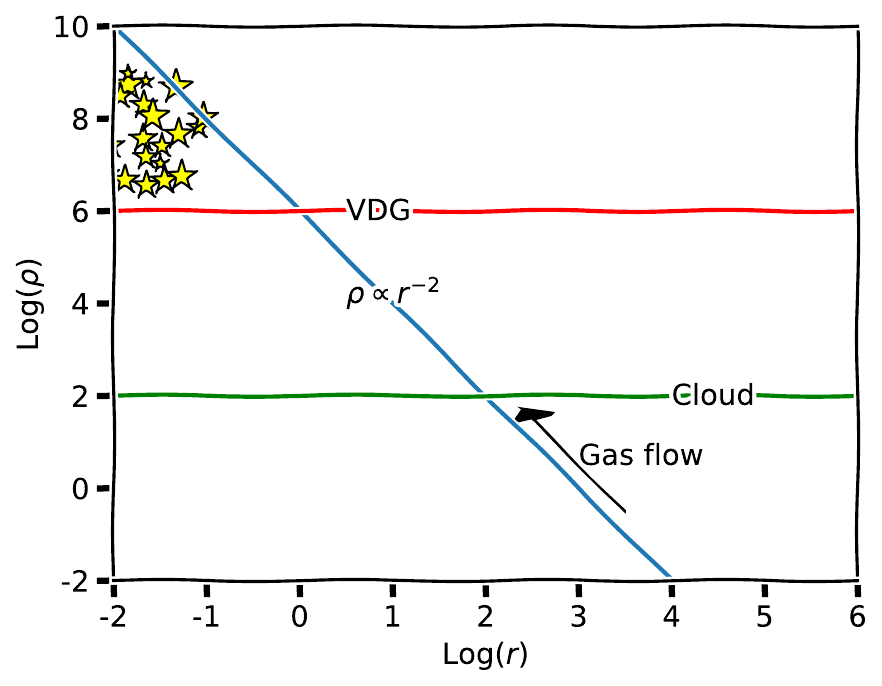}
\caption{
Schematic diagram illustrating a power law relationship between density and radius ($\rho \propto r^{-2}$;  blue line) in arbitrary units. Horizontal lines mark specific densities: the cloud density is shown in green and the density of very dense gas (VDG) in red. Star formation within VDG is represented by yellow stars.
}
\label{fig:cartoon}
\end{figure}

\subsubsection{The low measured values of $\eff$ as a consequence of the properties of a collapsing cloud}
\label{sec:low_eff} 

The results from Sec.\ \ref{sec:results} show that the clouds in our simulations do exhibit low and stationary values of $\eff$, in agreement with observational determinations, despite the facts that {\it a)} the gas parcels within them are undergoing gravitational collapse flow onto the corresponding potential well and {\it b)} the {\it final} SFE is significantly larger ($\sim 10\%$) than the measured values of $\eff$ ($\sim 1\%$). How can this happen? A solution to this apparent conundrum can be found from the definition of $\eff$ in the fundamental SF law, eq.\ \eqref{eq:eff}. In this equation, the SFR is formally given by
\begin{equation}
    \SFR \equiv \frac{d M_*} {dt} \approx \frac{\Delta M_*} {\Delta t},
    \label{eq:approx_SFR}
\end{equation}
so that $\eff$ can be written as
\begin{equation}
    \eff = \bigg(\frac{\tauff}{\Delta t} \bigg) \bigg(\frac{\Delta M_*}{\Mgas} \bigg).
    \label{eq:eff_prod_of_ratios_formal}
\end{equation}
{ The problem with this approach is that, in} studies based on YSO counting, the SFR is approximated by its average value over the lifetime of the YSOs, given by the mass in YSOs divided by their typical lifetime, as indicated by eq.\ \eqref{eq:mean_SFR}. Therefore, the {\it estimated} $\eff$ can be written as
\begin{equation}
    \eff \approx \bigg(\frac{\tauff}{\tausf} \bigg) \bigg(\frac{M_*}{\Mgas} \bigg).
    \label{eq:eff_prod_of_ratios_est}
\end{equation}
Such an approximation is valid { in general} if the SFR is approximately constant over that time period \citep{Dib+18}. We next analyze both ratios of this equation.

\bigskip
\noindent
a) {\it Ratio of the free-fall time to the lifetime of YSOs in systems with rapidly increasing SFRs.}

\medskip
In eq. \eqref{eq:eff_prod_of_ratios_est}, if the mass in YSOs and its time derivative, the SFR, are increasing nearly exponentially, as expected for free gravitational collapse, and verified in our simulations (Figs.~\ref{fig:efficiencies} and \ref{fig:Msinks}), then the time-averaging can produce a serious underestimation of the actual, current SFR. To see this, let us explicitly write the exponential growth of the mass in YSOs:
\begin{equation}
    M_*(t) = M_0 \exp\left({\frac{t}{\tausfr}}\right),
    \label{eq:exponential_SFR}
\end{equation}
where $M_0$ is some reference mass, $t$ is the time measured since the onset of star formation, and $\tausfr$ is the $e$-folding time for the mass in YSOs. 

This $e$-folding time can be related to the free-fall time by assuming it is given by the growth rate $\omega$ of the Jeans instability, for which the dispersion relation reads \citep[see, e.g., eq.\ (8.18) of] [] {Shu92}:
\begin{eqnarray}
    \omega^2 &=& k^2 c_{\rm s}^2 -4 \pi G \rho_0,\nonumber \\
             &=& \frac{2 \pi}{\tausfr},
\label{eq:Jeans_disp_rel}
\end{eqnarray}
where $k = 2 \pi/\lambda$ is the wavenumber of the applied perturbation and $\lambda$ is its wavelength, and the secon equality establishes the relationship between the growth rate and the characteristic timescale, which we identify with the $e$-folding time. The maximum growth rate occurs in the limit $k \rightarrow 0$, in which case the growth rate satisfies $i \omega = i 2 \pi/\tau = - \sqrt{4 \pi G \rho_{0}}$, and so the $e$-folding time is
\begin{equation}
    \tausfr = \left(\frac{\pi}{G\rho_{0}}\right)^{1/2}.
    \label{eq:tausfr}
\end{equation}
Comparing with the free-fall time, given by eq.\ \eqref{eq:tauff}, we see that $\tausfr \approx 3.27 \tauff$ { (or $\tauff/\tausfr \approx 0.31$)}. 

Fig. \ref{fig:Msinks} shows the evolution of sinks mass (black lines) for the HFS (left) and the FS (right panel). Red lines represent the best exponential fits:

  \begin{itemize}
      \item HFS:  $M_*(t) = (18.18 \, \Msun) \exp\left({\frac{t}{0.34 \, {\rm Myr}}}\right).$
      \item HF:  $M_*(t) = (0.23 \, \Msun) \exp\left({\frac{t}{0.61 \, {\rm Myr}}}\right).$
  \end{itemize}

\begin{figure*}
\includegraphics[width=0.45\textwidth]{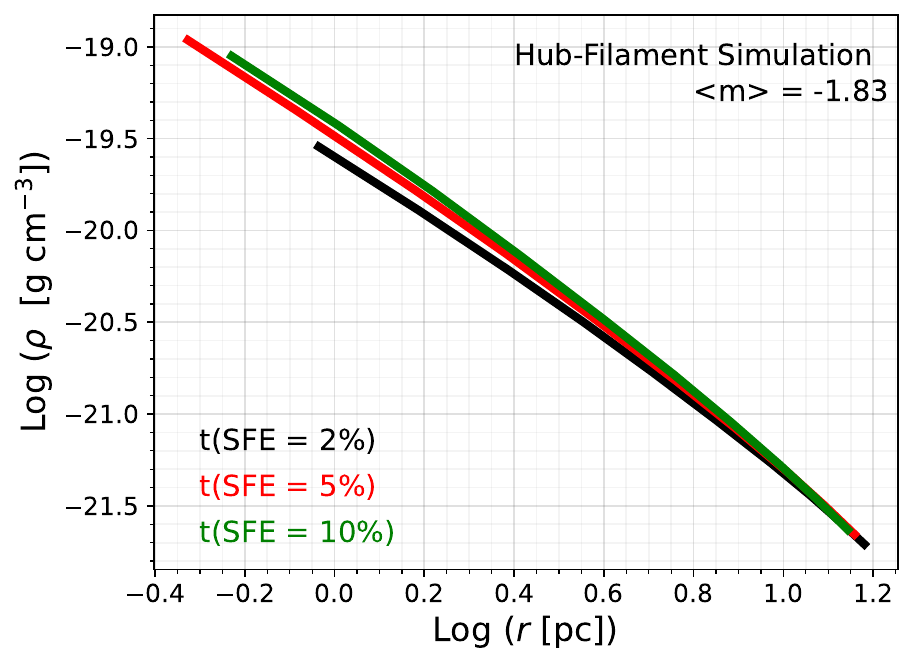}
\includegraphics[width=0.45\textwidth]{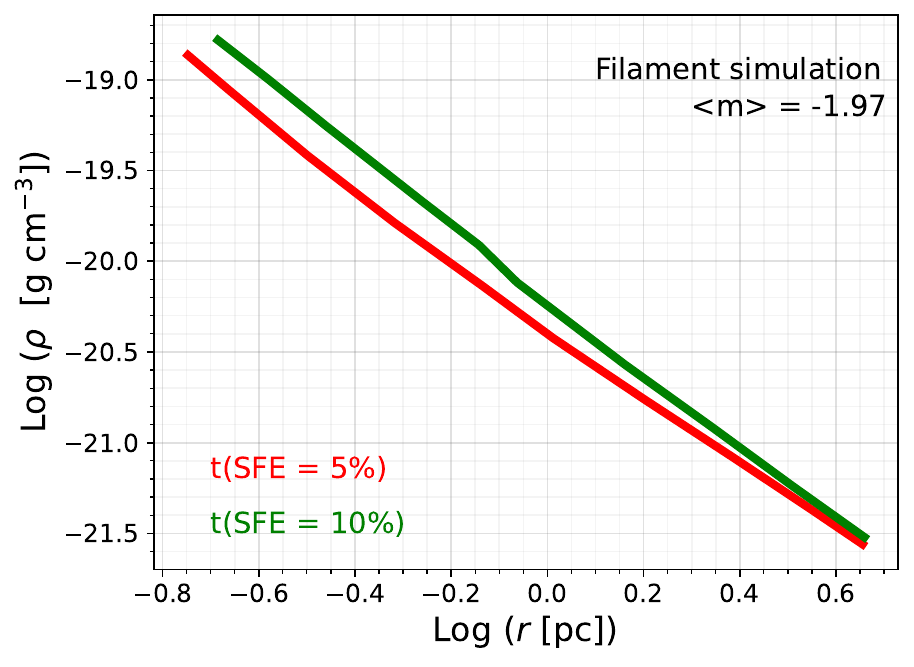}
\caption{Spherically-averaged density ($\rho$) as a function of radius ($R$) for the HFS (left) and FS (right) regions. The values of $R$ and $\rho$ were computed using Equations \eqref{eq:r} and  \eqref{eq:rho}. In both cases, $\rho$ follows a power-law dependence on $R$ with an exponent very close to $-2.0$. The average slope, $\langle m \rangle$, is calculated as the mean of the slopes derived from each individual line.
}
\label{fig:rho-r}
\end{figure*}

Therefore, we can see that the $e$-folding time is $\tausfr \approx 0.34$  Myr for run HFS and $\tausfr \approx \, 0.61$ Myr for the FS (during its exponential growth phase). This implies that the free-fall time corresponding to the instantaneous SFR  is $\tauff \sim$ {\blue 0.10} ~Myr for the HFS simulation and $\tauff \sim$ 0.18 Myr for the FS. Thus, the relevant free-fall time for the present-day SFR in the HFS is $\sim$ 5 $\times$ smaller (and $\sim$2.8 $\times$ smaller for the FS) than the typically assumed lifetime of embedded YSOs \citep[$\sim$0.5 Myr, e.g.,][]{Lada+13}, and $\sim$ 20 $\times$ smaller ($\sim$11.1 $\times$ smaller for the FS) than that of general YSOs \citep[$\sim$2 Myr, e.g.,][]{Evans+09}. In our simulations, we use $\tausf$ as an observational proxy for $\tausfr$, and the values of $\tausf$ used in our analysis for the HFS are $\sim$1.2, $\sim$1.5, and $\sim$1.7 Myr (see \S \ref{subsec:structure}). Consequently, $\tausf$ is $\sim$12, 15, and 17$\times$ larger than $\tauff$. Similarly, for the FS, $\tausf$ takes the values $\sim$3.4 and 4.2 Myr, resulting in a $\tausf$ that is $\sim$18 and 23 $\times$ larger than $\tauff$. Therefore, the first factor in Eq. \eqref{eq:eff_prod_of_ratios_est} is in general small, typically of the order $\sim 0.1$. Note that this argument applies mainly to the gas directly involved in the SF process, specifically within the inner contours. In the outer contours, $\tauff$ is on the order of magnitude of $\tausf$. However, while the mass in stars remains constant in the outer contours, the mass of the gas increases linearly with the contour size \citep[see][]{Ballesteros-Paredes+23}. Moreover, in our models the density exhibits a power-law dependence on the contour size, with an exponent close to $-2.0$ (see Fig. \ref{fig:rho-r}), implying a linear scaling between mass and size. This results in a net effect of maintaining a low $\epsilon_{\rm ff}$ (see below).

\bigskip
\noindent
b) {\it Ratio of stellar mass to gas mass.} 
\medskip

Let us now consider the mass ratio in eq.\ \eqref{eq:eff_prod_of_ratios_est}. For an exponentially growing YSO population, two conditions are satisfied: First, the mass in YSOs produced over the last $e$-folding time of the currently-star-forming very dense gas (VDG) is of the order of the total mass in YSOs produced over the full SF episode. Second, the mass in just-formed YSOs is of the order of  the gas mass of the VDG, since this is the gas mass that has been converted to stars in the last $e$-folding time. Therefore, we can write $M_* \sim ~\Msf$, 
where $\Msf$ is the mass of the VDG.

In addition, gravitational collapse, even when highly anisotropic and filamentary, develops a spherically-averaged radial density profile near $r^{-2}$. Indeed, \citet{Gomez_2021}
presented a compilation of reported density profiles for both low- and high-mass dense cores, finding mean logarithmic slopes $\sim -1.9$ for the former, and $\sim -1.7$ for the latter, and further showed that the slope of $-2$ is an attractor for the slope in the spherical collapse case. Similarly, for parsec-scale clumps, \citet{Peretto+23} also found typical spherically-averaged slopes $\sim -2$. Our own simulations exhibit a spherically-averaged density profile close to $r^{-2}$, as shown in Fig.\ \ref{fig:rho-r}. This behaviour can be understood from Poisson’s equation, which implies that the gravitational potential has a much smoother spatial distribution than the underlying density. This suggests that the dynamics of the spherical collapse problem provides a good approximation to the more complex, anisotropic case that develops in simulations. In the spherical case, a collapsing, self-gravitating density structure develops a close to $-2$ logarithmic slope \citep{Li+18, Gomez+21}, in agreement with the above quoted results.

Now, for a near $r^{-2}$ density profile, the mass, the free-fall time, and the $e$-folding time all scale linearly with radius. Therefore, the ratio $\Delta M_*/\Mgas$ in eq.\ \eqref{eq:eff_prod_of_ratios_formal} is of the order of $r_{\rm VDG}/r_{\rm cloud}$, where $r_{\rm VDG}$ is the size of the currently-star-forming VDG, and $r_{\rm cloud}$ is the size of the whole region where $\eff$ is being evaluated. Typically, this ratio is also very small (see Fig. \ref{fig:cartoon} for a schematic representation of this system).\\

For example, at the time when the SFE is $\sim$10\%, in the HFS simulation, we identify the contour with an average density closest to $n \sim 5 \times 10^4\, \ppcc$ and compute its mass to be $\sim$2436$\, \Msun$ (see Sec. \ref{subsec:structure} for details). This is $\sim 16 \times$ smaller than the total mass enclosed by the outer gas surface density contour. Similarly, in the FS, the mass within the contour closest to $n \sim 4 \times 10^3\, \ppcc$ is $\sim 506 \, \Msun$, which is $\sim 7 \times$ smaller than the total mass enclosed by the outer gas surface density contour.\\

We can conclude then that the small {\it measured} value of $\eff$ in YSO-counting studies follows from the rapidly accelerating nature of the SFR and the fact that the gas that is actually forming stars at any given time is much denser than the mean values in the regions where $\eff$ is being measured. First, the very large density of the star-forming gas implies that its corresponding free-fall time is much smaller than the YSO lifetime over which the SFR is averaged. Second, the exponential growth of the SFR causes that the total mass in YSOs is given essentially by the instantaneous mass in the very-high-density region that is currently forming stars. This, together with the near-$r^{-2}$ and nearly stationary density profile of the collapsing structure, implies that the stellar mass is always much smaller than the total mass in the clouds or cores whose $\eff$ is being measured. Thus, the two factors in eq.\ \eqref{eq:eff_prod_of_ratios_est} are small, causing the smallness of $\eff$ even in regions undergoing free collapse.

%%%%%%%%%%%%%%%%%%%%%%%%%%%%%%%%%%%%%%%%%%%%%%%%%%%%%%%%%%%%%%%%%%%
\bigskip
\subsubsection{The constancy of $\eff$ from gravitationally-driven accretion flow} \label{sec:const_eff}

The {\it constancy} of $\eff$ over both time and space also remains to be explained if the clouds are undergoing free gravitational contraction. Concerning the spatial dependence, we again note, as in the previous subsection, that, for a density profile near $r^{-2}$, both $\Mgas$ and $\tauff$ scale linearly with $r$, and therefore, from eq.\ \eqref{eq:eff_prod_of_ratios_formal}, $\eff$ is independent of radius. This can also be seen from eq.\ \eqref{eq:eff}, which, under the above consideration, shows that $\eff$ is proportional only to the instantaneous SFR. Since star formation only occurs at the density peaks, lower density isocontours around them that do not include any additional star-forming sites will have the same SFR, and thus $\eff$ will be independent of the isocontour's radius at any moment in time.

Moreover, the same argument applies to the temporal dependence. According to the similarity spherical collapse solution by \cite{Whitworth_Summers85}, at large radii, where the collapse is dominated by the gravity of the gas rather than that of the stars, the density profile is given by
\begin{equation} \label{eq:WS85}
\rho t^2 \propto \left(\frac{r}{c_{\rm s} t}\right)^{-2}.    
\end{equation} 
Therefore, the  density profile remains independent of time for gas undergoing this type of collapse flow \citep[see also] [] {Murray+15}. This again implies that the radial profiles of the mass and of the free-fall time are also time-independent, and therefore, so is $\eff$. That is, as proposed by \citet{Ballesteros-Paredes+23}, the reason for $\eff$, 
the ratio of SFR to gas-infall rate, to remain constant spatially and in time is the very collapse of the cloud. If some parts of the gas under consideration were not collapsing, $\eff$ would not be constant. This result can be thought of as a consequence of the fact that the rate at which the mass is transferred from one scale to the other under gravitational collapse is independent of radius for an object with an $r^{-2}$ density profile \citep{Li18, Gomez_2021}, and it remains so during the process of collapse, as would be expected for an intrinsically phenomenon such as gravity.

We can conclude then that the spatial and temporal constancy of $\eff$ is due to the fact that the two factors on the right-hand-side of eq.\ \eqref{eq:eff_prod_of_ratios_formal} vary inversely proportional to each other with radius for a collapse-generated density profile near $r^{-2}$, and this profile is approximately stationary over time. 

\subsection{The spatially intermittent nature of star formation and
  the density of clusters}
\label{sec:interm_cluster_dens} 

\begin{figure}
\includegraphics[width=\columnwidth]{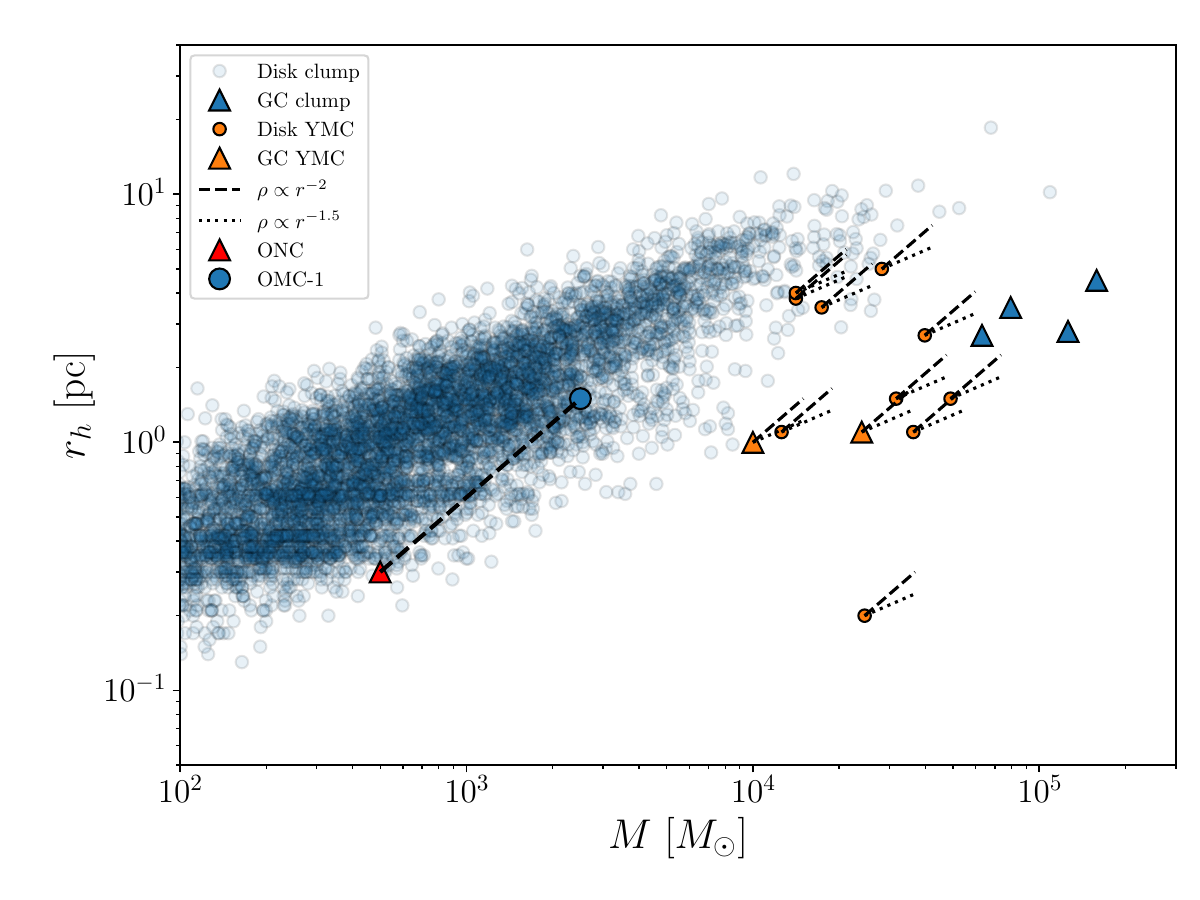}
\caption{Plot of radius versus mass for gas clumps in the Galactic disk (blue light circles) from ATLASGAL, clumps near the Galactic center (blue triangles), young massive clusters (YMCs; $M>10^4 \, \Msun$; orange circles), Galactic center YMCs (orange triangles), the Orion Nebula Cluster (ONC; red triangle) and its parent clump OMC-1 (blue circle). The black dashed and dotted lines represent the radius as a function of mass for density profiles $\rho \propto r^{-2}$ ($r \propto M$) and $\rho \propto r^{-3/2}$ ($r \propto M^{2/3}$), respectively. The black dashed line joining the ONC and OMC-1 has a slope of 1. Figure adapted from \citet[][Fig. 14, see references therein]{Krumholz+19}.}
\label{fig:cluster_dens}
\end{figure}

Another important implication of the spatially concentrated ({\it spatially intermittent}, in nonlinear dynamic terms) of the star formation process discussed in Sec. \ref{sec:low_eff}, is that star clusters appear to be significantly denser than gas clumps out of which they form \citep[e.g., Sec.\ 3.4.2 and Fig.\ 14 of] [] {Krumholz+19}. This higher density has been presented by those authors as a ``problem'' of the GHC scenario, but actually, \citet{BonillaBarroso+22} using numerical simulations \citep[e.g.,][]{Kuznetsova+15} shown that this is a natural outcome of the GHC model. They pointed out that during the process of gravitational collapse, stellar clusters tend to develop strong density concentrations. This occurs due to a combination of {\it i)} gas starvation because stars are forming in the central collapsing region and {\it ii)} the incorporation of newborn stars from the surrounding vicinity. Here, we further argue that this observation may be a consequence of:

\begin{enumerate}

\item the spatially intermittent nature of SF, 

\item the continuous gas inflow to the star-forming region, and

\item the exponentially increasing pace of the SFR,

\end{enumerate}
all of which are actually features of the GHC scenario, as dicussed above. In particular, item (i) is a consequence of item (ii); that is, the spatially intermittent nature of star formation is a direct consequence of the need for the gas to {\it flow} from low to high densities---i.e., from the density of the parent cloud to the density of the YSO itself. This means that the gas has to be collected from distant locations and moved into a very small region, where the density is very high. The continuous density-increasing flow manifests itself in the density profile of the collapsing gas, so that denser gas occupies smaller volume fractions. The gas forming stars at any given time (the VDG) has a density large enough that its free-fall time is much smaller than that of the parent cloud. Although the fragmentation in the region implies the appearance of multiple sites of collapse, these appear only in the VDG, not throughout the full volume of the cloud, whose lower-density parts are {\it flowing} towards the star forming sites.

The concentrated nature of the instantaneously-star-forming gas (the VDG) also has the implication that this mass is always a small fraction of the total cloud mass, since, as mentioned above, for a near $r^{-2}$ density profile, $M(r) \propto r$, so that the ratio of the VDG's mass ($M_{\rm VDG}$) to the cloud mass ($M_{\rm cl}$), satisfies
\begin{equation}
    \frac{M_{\rm VDG}}{M_{\rm cl}} = \frac{r_{\rm VDG}}{r_{\rm cl}} = \frac{{\tauff}_{\rm ,VDG}}{{\tauff}_{\rm ,cl}},
    \label{eq:MVDG_to_Mcl}
\end{equation}
where $r_{\rm VDG}$ and $r_{\rm cl}$ are respectively the sizes of the VDG and of its parent cloud, and ${\tauff}_{\rm ,VDG}$ and ${\tauff}_{\rm ,cl}$ are their corresponding free-fall times. Since the VDG is defined as having ${\tauff}_{\rm ,VDG} \ll {\tauff}_{\rm ,cl}$, both the mass and the size of the VDG are much smaller than those of their parent clouds, by a factor of the order of the ratio of their sizes.

Concerning item (iii) above, we have shown in Sec.\ \ref{sec:low_eff} and Fig.\ \ref{fig:Msinks} that the local SFR (denoted $\dot M_*$) in a star-forming region accelerates at a nearly exponential rate before feedback interrupts the process. As also mentioned in Sec.\ \ref{sec:low_eff}, the exponential growth of the stellar mass has two important implications: First, the current total YSO mass is comparable to the mass in YSOs formed over the last $e$-folding time of the VDG and, second, if all of the VDG's mass is converted to stars in one $e$-folding time, then the total YSO mass, $M_{\rm YSO}$, satisfies
\begin{equation}
    M_{\rm YSO} \sim M_{\rm VDG},
    \label{eq:MYSOsimMVDG}
\end{equation}
if the dense gas mass is continuously replenished by the accretion flow. Therefore, the mass density in the forming cluster is of the order of the mass density in the VDG, which, however, is much larger than the density of its parent cloud, by a factor $\sim \left(M_{\rm VDG}/M_{\rm cl}\right)^{-2}$. 

This is illustrated in Fig.\ \ref{fig:cluster_dens} \citep[adapted from Fig. 14 in][]{Krumholz+19}, which shows a size-mass plot for a collection of young massive clusters (orange triangles) and massive molecular clumps from the ATLASGAL survey \citep[blue light circles] [] {Urquhart+18}, as compiled by \citet{Krumholz+19}, and toward lower masses we show the Orion Nebula Cluster (ONC; red rectangle) and its parent clump, OMC-1 (blue circle), in the Orion Molecular Cloud, using the data compiled by \citet{Vazquez-Semadeni+10}. The black dashed lines indicate the $r \propto M$ scaling corresponding to a $\rho \propto r^{-2}$ density profile, while the black dotted lines represent $r \propto M^{2/3}$, corresponding to a density profile $\rho \propto r^{3/2}$. This figure shows that the ONC connects almost precisely with the OMC-1 clump along the $r \propto M$ line (red dashed line), being a factor $\sim 5$ smaller and less massive than the cloud and therefore a factor $\sim 25$ times denser. This suggests that the dashed lines should be extended to greater lengths and thus connect clusters with clumps that are typically $\sim 10$ times larger and more massive, and $\sim 100$ times less dense. This is also consistent with the typical final SFE observed in star-forming clouds, $\sim 10\%$, since the typical cluster mass is therefore $\sim 10\%$ of its parent clump's mass.

\subsection{Quenching the star formation}

In this paper, we have presented numerical simulations and analytical arguments concerning the establishment of the SK relation and the determination of $\eff$ only during the evoutionary stages dominated by infall; i.e. when feedback from locally-formed stars is negligible. Therefore, by design, our numerical simulations, which do not include feedback, are analyzed only up to moderate ($\lesssim 10\%$) {\it final} SFEs. Certainly, without the action of a mechanism that breaks up the cloud, our final efficiencies may grow to values substantially larger \citep[see, e.g., ][]{Vazquez-Semadeni+10, Colin+2013, Geen+2017, Li+2019, Grudic+2019, Kim+21, suin2023stellar}. Nevertheless, the results presented in this work without continuously driven turbulence nor stellar feedback imply that these agents are irrelevant in setting the star formation laws. Instead, the actual contribution of the stellar feedback must be shutting down the local process of collapse, by ionizing, heating up and eventually destroying the cloud, avoiding a large final efficiency of star formation. But in principle, there is no need for turbulence or feedback in settling the low values, spatial an temporal constancy of $\eff$. Thus, the large galactic depletion times of molecular gas are not due to star formation being "slow", but rather to it being extremely efficient in forming stars whose feedback destroys their parent cloud, preventing the present efficiency from increasing to large values.

\section{Summary and conclusions}
\label{sec:conclusions}

In this work, we have analysed two simulated star-forming structures that evolve according to the global hierarchical collapse scenario (GHC), a Hub-filament simulation (HFS) and a Filament simulation (FS) that extend over up to more than 10 pc. In this scenario, MCs and their substructures can be conceptualised as {\it sections} of a continuous flow in which the gas is passing from low- to high-density regions, increasing in mass and density. As they grow, these structures eventually become gravitationally unstable and start to form stars at an increasing rate. The entire lifecycle of the MC, from its formation as a cold neutral cloud until its eventual destruction or dispersal by newborn massive stars, extends over a few tens of Myrs. However, once star formation initiates, these structures display relatively short lifespans, of only a few {\it global} free-fall times \citep[see, e.g.,][and references therein]{Vazquez_Semadeni+19}.\footnote{This free-fall time is computed using the mean density of dense gas (with $n \ge 100 \, \ppcc$) at the moment when the first sink forms within the cloud.} 

We have measured the Schmidt-Kennicutt (SK) type relations and the ratio of the SFR to the gas-infall rate ($\eff$) parameter at three early times (when the SFE is $\sim$2, 5, and 10\%){ ,} following an observationally-motivated procedure: from surface density maps, we traced closed isocontours within which we measured the average surface gas density, area, enclosed mass, mass density (and free-fall time), and the star formation rate by dividing the mass in stars by a characteristic star formation timescale defined as the period elapsed between the formation of the first sink particle and the time under analysis.

In the analysed filaments, we recovered the observed intracloud SK-type relations measured in resolved galactic clouds: the classical, $\Sigmasfr \propto \Sigmamol^2$, and the more fundamental (tighter) relation, $\Sigmasfr \simeq \eff \Sigmamol/\tauff$ with no substantial variations among the analysed times. Even more, we also found consistency with the observations in the measured $\eff$. When calculating this parameter on various surface density isocontours, we obtain low values ($\eff \in [0.014, 0.063]$) with a distribution that is almost flat, which is consistent with recent resolved observations \citep[e.g.,][]{Pokhrel+21}.

The consistency of our results with observational determinations of the SK law and of $\eff$ suggests that our simulated filaments, whose evolution is mainly driven by self-gravity, capture the essential physics that determine these properties, in agreement with \cite{BP+23}, who claim that the low values of $\eff$ and its flat distribution in space are a consequence of collapse. In the dynamical GHC scenario, we propose the following explanations: {\it a)} The spatial and temporal stationarity of $\eff$ can be understood as a consequence of the fact that the gravitational collapse naturally develops a density slope near $\rho \propto r^{-2}$. This implies that the internal mass and the free-fall time of the structures both scale linearly with radius, \citep{Whitworth_Summers85, Gomez_2021}, causing $\eff$ to be time- and radius-independent (see eq. \ref{eq:WS85}); {\it b)} The low values of $\eff=(M_* / \Mgas)(\tauff / \tausf)$ (eq. \ref{eq:eff_prod_of_ratios_est}) arise from two key factors. First, the mass of young stars ($M_*$) is limited by the gas currently forming stars at sufficiently high densities. This high-density gas is a small fraction of the total cloud mass ($\Mgas$). Second, the very dense gas that is directly involved in the process of star formation has very short free-fall time compared with the YSO lifetime, causing the ratio $\tauff / \tausf$ to be small. Therefore, the two factors in the former equation that determine $\eff$ are small.

We also address the observation that star clusters are significantly denser than the gas clumps from which they originate, as discussed in \cite{Krumholz+19}. \cite{BonillaBarroso+22} show that this is a natural outcome of the GHC scenario and here we further propose that three key characteristics of collapsing clouds contribute to explain this observation: {\it i)} the spatially intermittent nature of star formation, {\it ii)} the continuous replenishment of gas to the star-forming regions, and {\it iii)} the rapid increase in the SFR.

We also emphasise that in the GHC scenario $\eff$ loses significance as an efficiency over one free fall time since clouds do not have a fixed mass and the SFR is not constant, and the quantities involved in the calculation of $\eff$ are not measured simultaneously: the free-fall time is estimated at the present time, while the SFR (or the depletion time) is averaged over some finite timescale, such as the lifetime of the YSOs. As the SFR increases, this average underestimates the instantaneous SFR. Therefore, $\eff$ is only meaningful as an efficiency for static clouds with constant total mass and SFR. These conditions are not satisfied in the case of clouds undergoing GHC, with increasing masses, densities and SFRs due to gravitational contraction. Consequently, interpreting $\eff$ as the ratio of two instantaneous rates (the SFR to the gas-infall rate) is a more accurate approach. Interestingly, our simulations produce surprisingly low and stable values of $\eff$ (0.04-0.063), despite achieving significantly higher final star formation efficiencies (up to 10\%). 

Our results disprove the often-made suggestion that models in the GHC scenario should have large values of $\eff$, and offer a new perspective on the role of self-gravity in defining the star formation laws. In addition, interpretations of $\eff$ in observational studies should be taken with caution, as $\eff$ alone may not suffice to distinguish between star formation scenarios. Instead, we suggest that the distinction may be provided by evidence of accelerating star formation, and that a more meaningful efficiency is the final one after a local SF episode is terminated.

\section*{Acknowledgements}
\addcontentsline{toc}{section}{Acknowledgements}

MZA acknowledges support from CONAHCYT grant number 320772.
The authors thankfully acknowledge computer resources, technical advice and support provided by: a) LANCAD-UNAM-DGTIC-188 and CONAHCYT on the supercomputer Miztli at DGTIC UNAM and; b) Laboratorio Nacional de Supercómputo del Sureste de México (LNS), a member of the CONAHCYT network of national laboratories. 
V.C. acknowledges funding from the National Science and Technology Council (NSTC 112-2636-M-003-001).
A.P. and E.V.-S.\ acknowledge ﬁnancial support from the UNAM-PAPIIT IG100223 grant. A.P. and J.B.P. acknowledge support from the CONAHCyT grant number 86372 of the `Ciencia de Frontera 2019’ program, entitled `Citlalc\'oatl: A multiscale study at the new frontier of the formation and early evolution of stars and planetary systems’. J.B.P. furthermore acknowledges UNAM-DGAPA-PAPIIT support through grant number {\tt IN-114422}
G.C.G. acknowledges support from UNAM-PAPIIT IN110824 grant.

%%%%%%%%%%%%%%%%%%%%%%%%%%%%%%%%%%%%%%%%%%%%%%%%%%

\section*{Data Availability}

The data generated for this article will be shared on request to the corresponding author.

%%%%%%%%%%%%%%%%%%%% REFERENCES %%%%%%%%%%%%%%%%%%

% The best way to enter references is to use BibTeX:

\bibliographystyle{mnras}

% \bibliography{refs} % if your bibtex file is called example.bib
\bibliography{refs, MisReferencias} % if your bibtex file is called example.bib

% Don't change these lines
\bsp	% typesetting comment
\label{lastpage}
\end{document}